\pdfoutput=1

\documentclass[11pt]{article}
\usepackage[utf8]{inputenc}
\usepackage[T1]{fontenc}
\usepackage[english]{babel}

\usepackage[totalwidth=480pt, totalheight=680pt]{geometry}

\usepackage{xcolor}
\colorlet{mygreen}{green!50!black}
\colorlet{myblue}{green!20!blue}

\usepackage{cite}
\usepackage{mathptmx}
\usepackage{amsmath,amsfonts,amsthm,amsbsy,amssymb}
\usepackage[hidelinks,colorlinks=true,linkcolor=myblue,citecolor=mygreen,urlcolor=myblue,bookmarksnumbered=true]{hyperref}
\usepackage{mathtools}
\usepackage{nicefrac}
\usepackage{mathrsfs}
\usepackage{bbm}
\numberwithin{equation}{section}
\usepackage{lmodern}
\usepackage[bottom]{footmisc}

\usepackage{tikz}
\usetikzlibrary{shapes.geometric}
\usepackage{floatrow}
\usepackage[labelfont=bf, font={small}]{caption}

\begin{document}
\begin{titlepage}
\setcounter{page}{0}
\begin{center}
\vspace*{1.5cm}
\textbf{\LARGE Coupling of Branes and Twisted Self-Duality}\\[2ex]
\textbf{\LARGE in the Maxwell-Chern-Simons Theory}\\
\vspace{2cm}
\textsc{\Large Hannes Malcha}\\
\vspace{1cm}
\textit{Max Planck Institute for Gravitational Physics (Albert Einstein Institute)}\\
\textit{D-14476 Potsdam, Germany}\\
\vspace{.5cm}
{\ttfamily \href{mailto:hannes.malcha@aei.mpg.de}{hannes.malcha@aei.mpg.de}}\\
\vspace{2cm}
\textbf{Abstract}
\end{center}
\small{
\noindent
We study three approaches to electric-magnetic duality in the 4-dimensional Maxwell theory coupled to a dyonic point charge and in the 5-dimensional Maxwell-Chern-Simons (MCS) theory coupled to an electric point charge and a magnetic string charge. The three approaches have been developed by Dirac, Bunster and Henneaux, and Pasti, Sorokin and Tonin (PST). In Dirac's formulation, the electric magnetic duality is realized only on the level of the equations of motion. The other two formulations introduce a dual (magnetic) gauge potential to induce manifest twisted self-duality in the action. In particular, we study the relations connecting the three approaches. The main results of this paper are the Bunster-Henneaux and PST formulations of the MCS theory with sources. We compare our result to the PST formulation of 11-dimensional supergravity coupled to the M2- and M5-brane by Bandos, Berkovits, and Sorokin.
}
\vspace{\fill}
\end{titlepage}

\section{Introduction}
The duality between electric and magnetic fields in the source-free Maxwell equations is one of the most profound symmetries in physics. Its original discovery is said to predate even the introduction of Maxwell's equations themselves. Over the years the electric-magnetic duality showed a remarkable resilience to countless generalizations and today it is ubiquitous in our formulations of string theory and supergravity.

Arguably the most important generalization of the electric-magnetic duality in the source-free Maxwell equations was the introduction of the magnetic monopole by Dirac \cite{Dirac:1931kp, Dirac:1948um}. He showed that the existence of only a single magnetic monopole in the universe can explain the quantization of all electric charges. However, this result comes with a literal string attached. By Gauss's law for magnetism, the flux of a magnetic field through any closed surface must vanish. However, if there is a magnetic pole inside the closed surface this is not true. Hence, the electromagnetic potential must be singular somewhere on the surface. Since this is true for any closed surface containing the magnetic monopole the electromagnetic potential is singular on a line of points extending outwards from the monopole, usually called the Dirac string.

The Dirac string is an unphysical object, which must be invisible to the charges placed in the system and cannot have any dynamic of its own. The former condition yields the famous Dirac quantization condition. If we assume so-called Dirac vetos, which will be explained in detail in the main text, the latter condition must be satisfied automatically in a Lagrangian formulation of the system. In \cite{Dirac:1948um} Dirac developed a Lagrangian theory that gives rise to duality symmetric Maxwell equations containing both electric and magnetic charges. 

However, even without sources the electric-magnetic duality symmetry does, a priori, not extend from the equations of motion to the action. To see this consider the source-free Maxwell equations in their covariant formulation
\begin{gather}
\begin{gathered}
\partial_\nu F^{(1) \, \mu\nu} = 0 \, ,\\
\partial_\nu (\star F^{(1) \, \mu\nu}) = 0 \, .
\end{gathered}
\end{gather}
Clearly these equations are invariant under the exchange of $F_{\mu\nu}^{(1)}$ and $\star F_{\mu\nu}^{(1)}$. However, in a Lagrangian formulation of Maxwell's theory the field strength $F_{\mu\nu}^{(1)}$ is merely a derived quantity stemming from an electromagnetic gauge potential $A_\mu^{(1)}$ such that $F_{\mu\nu}^{(1)} = 2 \partial_{[\mu} A_{\nu]}^{(1)}$. Once the potential is introduced the second equation becomes an identity whereas the first one remains an equation of motion. Hence one cannot rotate $F_{\mu\nu}^{(1)}$ into $\star F_{\mu\nu}^{(1)}$ off-shell. 

The canonical way to make the electric-magnetic duality transformations manifest in the action is to introduce a dual gauge potential $A_\mu^{(2)}$ and an associated dual field strength $F_{\mu\nu}^{(2)}$ such that the Bianchi identity of $F_{\mu\nu}^{(2)}$ is the equation of motion for $A_\mu^{(1)}$ and vice versa. Hence the field strengths $F_{\mu\nu}^{(1,2)}$ are duals of each other. This is known as twisted self-duality \cite{Cremmer:1998px}. In this paper, we discuss the two approaches to twisted self-duality developed by Bunster and Henneaux \cite{Bunster:2011qp} as well as by Pasti, Sorokin and Tonin (PST) \cite{Pasti:1995tn, Pasti:1996vs}.

Briefly speaking the idea of the Bunster-Henneaux approach to twisted self-duality is to introduce the spatial components of a dual gauge potential as the solution to the Gauss constraint. This eliminates the time component of the original gauge potential from the usual Dirac action \cite{Deser:1997mz}. The cost of this approach is the loss of manifest Lorentz invariance in the theory. However, the resulting action remains gauge invariant and becomes invariant under the exchange of $A_i^{(1)}$ and $A_i^{(2)}$. Making use of the Dirac string the generalization to included sources is immediate \cite{Deser:1997mz}.

\begin{figure}[t]
\begin{tikzpicture}
\draw [very thick] (5,0) ellipse (2 and 1);
\draw [very thick] (0,-5) ellipse (2 and 1);
\draw [very thick](10,-5) ellipse (2 and 1);
\node at (5,0) [color=black] {\textbf{\textsf{PST}}};
\node at (0,-5) [color=black] {\textbf{\textsf{Dirac}}};
\node at (10,-5) [color=black] {\textbf{\textsf{Bunster-Henneaux}}};
\draw[->, very thick, bend right=17] (2.5,-5.5) to (7.5,-5.5);
\draw[->, very thick, bend right=17] (7.5,-4.5) to (2.5,-4.5);
\draw[->, very thick] (7,-1) to (9,-3.5);
\draw[->, very thick] (3,-1) to (1,-3.5);
\node at (0,-2.25) {integrate out $A^{(2)}$};
\node at (5,-3.5) {integrate out $A^{(2)}$};
\node at (5,-6.5) {introduce $A^{(2)}$ as};
\node at (5,-7) {solution to Gauss constraint};
\node at (10,-2.25) {$v_\mu = (1,0,\ldots,0)$};
\end{tikzpicture}
\vspace{.5cm}
\caption{Schematic relation of the PST, Dirac, and Bunster-Henneaux actions. There is no systematic way to reach the PST action and it must always be verified by its relation to the other two actions. Integrating out $A^{(2)}$ is plugging the solution for the equation of motion of $A^{(2)}$ into the action, which eliminates $A^{(2)}$.}
\label{fig:diag}
\end{figure}
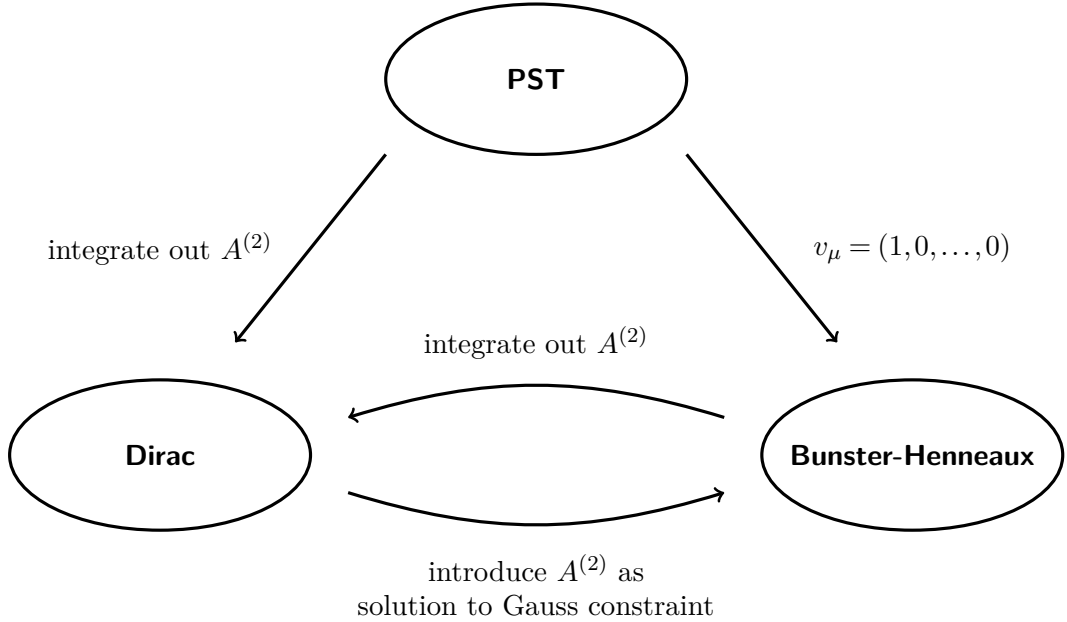

The PST approach on the other hand retains manifest Lorentz and gauge invariance by introducing an auxiliary field $a(x)$ which enters the action through the timelike harmonic unit vector
\begin{align}\label{eq:v}
v_\mu(x) \coloneqq \frac{\partial_\mu a(x)}{\sqrt{- (\partial^\nu a \, \partial_\nu a)}} \, , \quad v_\mu v^\mu = -1 \, .
\end{align}
Furthermore, the auxiliary field gives rise to additional gauge symmetries. When the auxiliary vector $v_\mu(x)$ is gauge fixed to have only a non-zero time component, \emph{i.e.} $v_\mu = (1, 0, \ldots, 0)$, the PST action agrees with the Bunster-Henneaux action. Integrating out the dual gauge potential $A_\mu^{(2)}$ from either the Bunster-Henneaux or the PST action reduces them to the Dirac action. These relations are summarized in figure \ref{fig:diag}. 

There are of course many more approaches to manifest electric-magnetic duality besides the ones discussed here. For example Sen \cite{Sen:2015nph, Sen:2019qit} and Mkrtchyan et al. \cite{Mkrtchyan:2019opf, Bansal:2021bis, Avetisyan:2021heg, Avetisyan:2022zza} proposed two different duality-symmetric formulations of type IIB supergravity and p-form theories respectively which are polynomial, Lorentz invariant and contain only a finite number of fields (see also \cite{Evnin:2022kqn} for a brief review). In particular, the Mkrtchyan et al. formulation has a further advantage over the PST formulation in that it can be applied to theories with arbitrary self-interactions in any dimension. Moreover, also theories with Chern-Simons interaction terms have been studied in this approach \cite{Evnin:2023ypu}. However, further work is needed to extend both the Sen and Mkrtchyan et al. formulations to include the coupling to external electric and magnetic charges.

As mentioned in the beginning electric-magnetic duality is present in a plethora of theories. In this paper, we focus on two of them. Namely, the 4-dimensional Maxwell theory coupled to a dyonic point charge as well as the 5-dimensional Maxwell-Chern-Simons (MCS) theory coupled to an electric point charge and a magnetic string charge. We provide a systematic study of the Dirac, Bunster-Henneaux, and PST approaches to electric magnetic duality for both theories. Particular emphasis is put on highlighting the relations of figure \ref{fig:diag}. The 5-dimensional MCS theory comes with a cubic Chern-Simons interaction term and corresponds to the bosonic sector of 5-dimensional supergravity in a fixed gravitational background \cite{Chamseddine:1980mpx}. Its generalization to 11 dimensions with a 3-form gauge potential describes the bosonic sector of flat 11-dimensional supergravity \cite{Cremmer:1978km}. 

The main results of this work are the PST and Bunster-Henneaux formulations of the 5-dimensional MCS theory coupled to electric and magnetic charges. We show that these formulations consistently reduce to the Dirac formulation of \cite{Bekaert:2002eq}. Moreover, we argue that this is not the case for a similar 11-dimensional PST formulation given in \cite{Bandos:1997gd}. Upon reducing it to 11-dimensional supergravity coupled to the electric M2- and the magnetic M5-brane, the Dirac 6-brane coupled to the M5-brane becomes dynamical, thus contradicting its unphysical nature. We discuss a solution to this issue based on our PST formulation of the 5-dimensional MCS theory.

The rest of this paper is organized as follows. In section \ref{sec:m} we consider the 4-dimensional Maxwell theory coupled to a dyonic point charge. We briefly review its Dirac, Bunster-Henneaux, and PST formulation as well as their relations to each other. In section \ref{sec:mcs} we repeat the analysis of section \ref{sec:m} but this time for the 5-dimensional MCS theory. Finally, section \ref{sec:conc} contains some brief concluding remarks. 

\section{Maxwell Theory}\label{sec:m}
In this section, we discuss electric-magnetic duality in the Maxwell theory. We work in 4-dimensional flat Minkowski space with the mostly plus metric $\eta_{\mu\nu} = (-, +, +, +)$. The totally anti-symmetric tensor is $\varepsilon_{0123} = 1$. We consider a point particle of mass $m$ with both electric charge $e$ and magnetic charge $g$ (\emph{i.e.} a dyon \cite{Schwinger:1968rq, Schwinger:1969ib, Zwanziger:1968rs}) moving through an electromagnetic field. The world-line of the point particle is parametrized by $z \equiv z(\sigma_1)$ as a function of the particle's proper time $-\infty \le \sigma_1 \le \infty$. The motion of the dyon is given by Lorentz's equation
\begin{align}\label{eq:m_eom_z}
m \ddot{z}_\mu &= \left( e F_{\mu\nu}^{(1)}(z) - g (\star F_{\mu\nu}^{(1)})(z) \right) \dot{z}^\nu \, .
\end{align}
The relative minus sign is to ensure the invariance of the equation under electric-magnetic duality transformations as we will see explicitly at the end of section \ref{sec:m_bh}. A dyon gives rise to both an electric current density $J_\mu^{(1)}$ as well as a magnetic current density $J_\mu^{(2)}$. Hence the appropriate Maxwell equations to describe the system are
\begin{gather}\label{eq:m_maxwell}
\begin{gathered}
\partial_\nu F^{(1) \, \mu\nu} = J^{(1) \, \mu} \, , \\
\partial_\nu (\star F^{(1) \, \mu\nu}) = - J^{(2) \, \mu} \, .
\end{gathered}
\end{gather}
The goal of this section is to formulate Dirac, Bunster-Henneaux, and PST actions whose equations of motion imply \eqref{eq:m_eom_z} - \eqref{eq:m_maxwell}. Furthermore, we will study the relations given in figure \ref{fig:diag}.

To write the actions we must introduce two vector fields $A_\mu^{(1,2)}$ associated to the field strengths
\begin{align}
F_{\mu\nu}^{(1,2)} \coloneqq 2 \partial_{[\mu} A_{\nu]}^{(1,2)} \, .
\end{align}
We may express the electric and magnetic currents as
\begin{align}\label{eq:m_J}
J_\mu^{(1,2)}(x) \coloneqq q^{(1,2)} \int \mathrm{d}z_\mu \ \delta^{(4)}(x-z)
\end{align}
with $q^{(1)} = e$ and $q^{(2)} = g$. To the dyon, we attach an electric or a magnetic Dirac string $G_{\mu\nu}^{(1,2)}$, which sweeps out a 2-dimensional worldsheet in spacetime. The sheet is parametrized by $y \equiv y(\sigma_1, \sigma_2)$ with $-\infty \le \sigma_1 \le \infty$ and $0 \le \sigma_2 \le \infty$. Moreover we set $y(\sigma_1, 0) \coloneqq z(\sigma_1)$. Thus the worldline of the current is the boundary of the worldsheet of the string. This implies that the $y_\mu(\sigma_1,\sigma_2)$ are dynamical variables just as the $z_\mu(\sigma_1)$. The Dirac string is defined as
\begin{gather}\label{eq:m_G}
\star G_{\mu\nu}^{(1,2)} = \frac{1}{2} \varepsilon_{\mu\nu\rho\lambda} \, G^{(1,2) \, \rho\lambda} \, , \\
G_{\mu\nu}^{(1,2)}(x) \coloneqq q^{(1,2)} \int \mathrm{d}y_\mu \wedge \mathrm{d}y_\nu \ \delta^{(4)}(x-y) \, .
\end{gather}
The definition and parametrization imply
\begin{align}
\partial^\mu G_{\mu\nu}^{(1,2)}(x) = - J_\nu^{(1,2)}(x) \, .
\end{align}
Finally, we define the extended field strengths
\begin{align}\label{eq:m_H}
H_{\mu\nu}^{(1,2)} \coloneqq F_{\mu\nu}^{(1,2)} + \star G_{\mu\nu}^{(2,1)} \, .
\end{align}
It is easy to see that with the above definitions
\begin{align}\label{eq:m_Bianchi}
\partial_\nu (\star H^{(1,2) \, \mu\nu}) = - J^{(2,1) \, \mu} \, .
\end{align}
When a Dirac string is part of an action, the variation with respect to its string variables $y_\mu(\sigma_1,\sigma_2)$ gives rise to an equation of motion similar to \eqref{eq:m_eom_z}. However, since the Dirac string is an unphysical object, unobservable in nature, it is crucial, that this additional equation of motion is automatically satisfied by any solution of the other equations of motion. Otherwise, the Dirac string becomes a dynamical object. In this paper, we will explicitly check the string equations of motion. 

\subsection{Dirac}
The action proposed by Dirac, describing the coupling of a dyon to non-linear electrodynamics, reads \cite{Dirac:1948um}
\begin{align}\label{eq:m_sd}
S_\mathrm{D} = - \frac{1}{4} \int \mathrm{d}^4x \ H_{\mu\nu}^{(1)} H^{(1) \, \mu\nu} + I_e + I_p \, , 
\end{align}
with
\begin{gather}
I_e = e \int \mathrm{d}z^\mu \ A_\mu^{(1)}(z) \, , \\
I_p = m \int \mathrm{d}\sigma_1 \ \sqrt{- \frac{\partial z_\mu}{\partial \sigma_1} \frac{\partial z^\mu}{\partial \sigma_1}} \, .
\label{eq:m_d_Ip}
\end{gather}
$I_e$ describes the interaction of the electric charge with the vector field. The interaction of the magnetic charge with the vector field is hidden in the Maxwell term through the definition of extended field strength \eqref{eq:m_H}. Thus, we see that the electric and magnetic charges appear asymmetrically in this action. In particular, the action is not duality invariant. Finally, $I_p$ is the standard action of only the dyon.

To obtain the equations of motion we extremize the action \eqref{eq:m_sd} with respect to $A_\mu^{(1)}$, $y_\mu$ and $z_\mu$. This will give us an equation of motion of the field, the dyon, and the Dirac string attached to the dyon. The details of the extremization with respect to $y_\mu$ and $z_\mu$ are given in \cite{Dirac:1948um}. The respective calculations are straightforward. For the vector field $A_\mu^{(1)}$ we obtain the equation of motion
\begin{align}\label{eq:m_d_eom}
\partial_\nu H^{(1) \, \mu\nu}(x) = J^{(1) \, \mu}(x) \, .
\end{align}
Moreover, the extended field strength satisfies the Bianchi identity \eqref{eq:m_Bianchi}.
\begin{align}\label{eq:m_d_bianchi}
\partial_\nu (\star H^{(1) \, \mu\nu})(x) = - J^{(2) \, \mu}(x) \, .
\end{align}
From extremizing the action with respect to $z_\mu$ we obtain the equation of motion of the point charge
\begin{align}\label{eq:m_d_eom_z}
m \ddot{z}_\mu &= \left( e F_{\mu\nu}^{(1)}(z) - g (\star H_{\mu\nu}^{(1)})(z) \right) \dot{z}^\nu \, ,
\end{align}
Finally, the equation of motion of the Dirac string coming from the extremization with respect to $y_\mu$ is
\begin{align}
\partial_{[\mu} \left(\star H_{\nu\rho]}^{(1)}(y)\right) = 0 \, ,
\end{align}
which is equivalent to
\begin{align}\label{eq:m_d_eom_y}
\partial_\nu H^{(1) \, \mu\nu}(y) = 0 \, .
\end{align}
We see that this equation agrees with the field equation of motion \eqref{eq:m_d_eom} provided that $J_\mu^{(1)}(y) = 0$, \emph{i.e.} if the string never passes through the charged particle. This additional assumption is known as the Dirac veto \cite{Dirac:1948um}. For a more general action of $n$ dyons and $n$ Dirac strings the Dirac veto states that none of the particles lies on any of the strings. In \eqref{eq:m_d_eom_z} the Dirac veto implies $G_{\mu\nu}^{(2)}(z) = 0$, \emph{i.e.} that $\star H_{\mu\nu}^{(1)}(z) $ must be replaced by $\star F_{\mu\nu}^{(1)}(z)$. Then we see that \eqref{eq:m_d_eom_z} agrees with \eqref{eq:m_eom_z}. The field equation of motion \eqref{eq:m_d_eom} and the Bianchi identity \eqref{eq:m_d_bianchi} are the Maxwell equations \eqref{eq:m_maxwell} with $F_{\mu\nu}^{(1)}$ replaced by the extended field strengths $H_{\mu\nu}^{(1)}$. This makes the second Maxwell equation compatible with the introduction of the gauge potential $A_\mu^{(1)}$. 

Since $\star \star H_{\mu\nu}^{(1)} = - H_{\mu\nu}^{(1)}$, the field equation of motion \eqref{eq:m_d_eom} and the Bianchi identity \eqref{eq:m_d_bianchi} are invariant under $\mathrm{SO}(2)$ transformations of the field strengths and current densities
\begin{align}\label{eq:m_d_so2}
\begin{pmatrix} H_{\mu\nu}^{(1) \, \prime} \\ \star H_{\mu\nu}^{(1) \, \prime} \end{pmatrix} = 
\begin{pmatrix} \cos \xi & \sin \xi \\ - \sin \xi & \cos \xi \end{pmatrix} \begin{pmatrix} H_{\mu\nu}^{(1)} \\ \star H_{\mu\nu}^{(1)} \end{pmatrix} \, , 
\quad \quad \begin{pmatrix} J_\mu^{(2) \, \prime} \\ J_\mu^{(1) \, \prime} \end{pmatrix} = 
\begin{pmatrix} \cos \xi & \sin \xi \\ - \sin \xi & \cos \xi \end{pmatrix} \begin{pmatrix} J_\mu^{(2)} \\ J_\mu^{(1)} \end{pmatrix} \, .
\end{align}
For this to be a true symmetry of the theory, the transformations should also leave the action invariant. Since the field strength and the current densities are derived quantities one must first define the transformations on the vector field $A_\mu^{(1)}$ and the charges $e$ and $g$. However, once the vector field is introduced \eqref{eq:m_d_bianchi} becomes an identity while \eqref{eq:m_d_eom} remains an equation of motion. Thus an action containing only $A_\mu^{(1)}$ cannot be duality symmetric.

The Bunster-Henneaux and PST formulations discussed below introduce a dual gauge potential $A_\mu^{(2)}$ via its extended dual field strength $H_{\mu\nu}^{(2)}$ into the action \eqref{eq:m_sd} such that the equation of motion for $A_\mu^{(2)}$ is the Bianchi identity of $H_{\mu\nu}^{(1)}$ and vice versa. The duality transformations \eqref{eq:m_d_so2} can then be defined directly on the fields $A_\mu^{(1,2)}$ and charges $q^{(2,1)}$ and they leave the actions invariant.

\subsection{Bunster-Henneaux}\label{sec:m_bh}
In this subsection, we briefly review the Bunster-Henneaux action for the 4-dimensional Maxwell theory with electric and magnetic charges. We will sacrifice manifest Lorentz invariance to obtain a two-potential formulation of \eqref{eq:m_sd} that treats electric and magnetic sources on the same footing. The results presented here were first obtained by Bunster, Deser, Gomberoff, and Henneaux in \cite{Deser:1997se, Deser:1997mz}. 

Let $i,j,\ldots \in \{1,2,3 \}$ and perform the standard Legendre transformation on the spatial components of the vector potential $A_i^{(1)}$ such that \eqref{eq:m_sd} becomes
\begin{align}\label{eq:m_sbh_1}
S_\mathrm{BH} &= \int \mathrm{d}^4x \ \left[ \dot{A}_i^{(1)} \pi^i + \star G_{0i}^{(2)} \pi^i - \mathcal{H}(\pi,B) + (\partial_i \pi^i + J^{(1) \, 0}) A_0^{(1)} \right] + I_e^{\prime} + I_p \, ,
\end{align}
where be $\pi^i = - H^{(1) \, 0i}$ is the canonical momentum and $B$ is the magnetic field, to be defined below. $I_p$ is as before and
\begin{gather}
I_e^\prime = e \int \mathrm{d}z^i \ A_i^{(1)}(z) \, , \\
\mathcal{H}(\pi,B) = H_{0i}^{(1)} \pi^i + \frac{1}{4} H_{\mu\nu}^{(1)} H^{(1) \, \mu\nu} \, .
\end{gather}
We introduce the dual vector potential $A_i^{(2)}$ as a solution to the Gauss constraint in \eqref{eq:m_sbh_1} by defining
\begin{align}
- \pi^{(1) \, i} \coloneqq - \varepsilon^{0ijk} \partial_j A_k^{(2)} + G^{(1) \, 0i} \, , 
\end{align}
such that $\partial_i \pi^{(1) \, i} = - J^{(1) \, 0}$. This choice breaks the manifest Lorentz invariance of the action \eqref{eq:m_sbh_1}. Furthermore, let\footnote{Notice that the definition of $\mathbf{q}^a$ is reversed compared to the definition of $q^{(1,2)}$ in the previous subsection.}
\begin{align}
\begin{aligned}
&\mathbf{q}^a \coloneqq (g,e) \, , \\
&\mathbf{A}^a \coloneqq (\mathbf{A}^{(1)}, \mathbf{A}^{(2)}) \, , \\
&\mathbf{B}^a \coloneqq (\mathbf{B}, - \boldsymbol{\pi}) 
\end{aligned}
\end{align}
with
\begin{align}\label{eq:m_bh_b}
\mathbf{B}^a &= - \nabla \times \mathbf{A}^a + \boldsymbol{\beta}^a \, 
\end{align}
and
\begin{align}
\boldsymbol{\beta}^1 \coloneqq G^{(2) \, 0i} \, , 
\quad 
\boldsymbol{\beta}^2 \coloneqq G^{(1) \, 0i} \, .
\end{align}
Compared to \cite{Deser:1997se} we have an extra minus sign in the definition of $\mathbf{B}^a$. This is because we work with $\varepsilon^{0ijk}$ instead of $\varepsilon^{ijk}$. Moreover, we define
\begin{align}\label{eq:m_bh_a}
\alpha_i^1 \coloneqq \frac{1}{2} \varepsilon_{0ijk} G^{(2) \, jk} \, ,
\quad
\alpha_i^2 \coloneqq \frac{1}{2} \varepsilon_{0ijk} G^{(1) \, jk} \, ,
\end{align}
such that the action \eqref{eq:m_sbh_1} becomes
\begin{align}\label{eq:m_sbh_2}
S_\mathrm{BH} &= \int \mathrm{d}^4x \ \left[ - \mathbf{B}^2 \cdot (\dot{\mathbf{A}}^1 + \boldsymbol{\alpha}^1) - \mathcal{H}(\mathbf{B}) \right] + I_e^\prime + I_p \, .
\end{align}
We can then put everything into a manifestly symmetric formulation, by noticing the (symmetric) identity
\begin{align}\label{eq:m_bh_aux_1}
\frac{1}{2} \int \mathrm{d}^4x \ k_{ab} \, \mathbf{B}^a \cdot (\dot{\mathbf{A}}^b + \boldsymbol{\alpha}^b) - \frac{1}{2} k_{ab} \, \mathbf{q}^a \int \mathrm{d}\mathbf{z} \cdot \mathbf{A}^b(z) = \int \mathrm{d}^4x \ \left[ \partial_\mu V^\mu + \frac{1}{2} \, k_{ab} \, \boldsymbol{\alpha}^a \cdot \boldsymbol{\beta}^b \right] \, ,
\end{align}
with some 1-form $V^\mu$. Here $k_{ab}$ is the symmetric matrix
\begin{align}
k_{ab} = \begin{pmatrix} 0 & 1 \\ 1 & 0 \end{pmatrix} \, .
\end{align}
It can be shown that the entire right-hand side of \eqref{eq:m_bh_aux_1} has vanishing variational derivative, for both the variation with respect to the vector fields as well as the string coordinates. Thus, we can add the left hand side of \eqref{eq:m_bh_aux_1} to the action \eqref{eq:m_sbh_2} without changing the equations of motion and obtain
\begin{align}\label{eq:m_sbh_3}
S_\mathrm{can} = \int \mathrm{d}^4x \ \left[ \frac{1}{2} \epsilon_{ab} \, \mathbf{B}^a \cdot (\dot{\mathbf{A}}^b + \boldsymbol{\alpha}^b) - \mathcal{H} \right] + I_q^\prime + I_p \, ,
\end{align}
with $\epsilon_{ab}$ anti-symmetric and
\begin{align}\label{eq:m_bh_Iq}
I_q^\prime = - \frac{1}{2} \epsilon_{ab} \, \mathbf{q}^a \int \mathrm{d}\mathbf{z} \cdot \mathbf{A}^b(z) \, .
\end{align}
The Hamiltonian can easily be computed and reads
\begin{align}
\mathcal{H} = \frac{1}{2} \left( \mathbf{B}^1 \cdot \mathbf{B}^1 + \mathbf{B}^2 \cdot \mathbf{B}^2 \right) \, .
\end{align}
We see that the action \eqref{eq:m_sbh_3} is off-shell invariant under $\mathrm{SO}(2)$ rotations of the vector potentials $\mathbf{A}^a$ and charges $\mathbf{q}^a$. We proceed to compute the equations of motion. Extremizing the action \eqref{eq:m_sbh_3} with respect to $A_i^{(1)}$ yields
\begin{align}\label{eq:m_bh_eom_1}
0 = \partial_j \left( \varepsilon^{0ijk} \dot{A}_k^{(2)} - G^{(1) \, jk} + H^{(1) \, ij} \right) \, .
\end{align}
This equation is solved locally by introducing a new variable $A_0^{(2)}$ such that
\begin{align}\label{eq:m_bh_eom_sol_1}
\varepsilon^{0ijk} \dot{A}_k^{(2)} - G^{(1) \, jk} + H^{(1) \, ij} = \varepsilon^{0ijk} \partial_k A_0^{(2)} \, .
\end{align}
Rewriting this formula leads to
\begin{align}
H^{(1) \, ij} = - \varepsilon^{ij0k} H_{0k}^{(2)} \, .
\end{align}
Similarly the equation of motion for $A_i^{(2)}$ is 
\begin{align}\label{eq:m_bh_eom_2}
0 = \partial_j \left( \varepsilon^{0ijk} \dot{A}_k^{(1)} - G^{(2) \, jk} - H^{(2) \, ij} \right) \, .
\end{align}
This is solved by re-introducing $A_0^{(1)}$ similar to \eqref{eq:m_bh_eom_sol_1}. Together the solutions of \eqref{eq:m_bh_eom_1} and \eqref{eq:m_bh_eom_2} are equivalent to
\begin{align}\label{eq:m_bh_sol}
H_{\mu\nu}^{(2)} = \star H_{\mu\nu}^{(1)} \, .
\end{align}
Taking the exterior derivative of \eqref{eq:m_bh_sol} and its dual we obtain the Maxwell equations \eqref{eq:m_maxwell}, \emph{i.e.}
\begin{align}
0 = \partial_\nu \left( H^{(1,2) \, \mu\nu} \pm \star H^{(2,1) \, \mu\nu} \right) = \partial_\nu H^{(1,2) \, \mu\nu} \mp J^{(1,2) \, \mu} \, .
\end{align}
Extremizing the action with respect to $z^0$ and $z^i$ we obtain
\begin{align}
 m \ddot{z}_0 = \left( - \frac{e}{2} \varepsilon_{0ijk} H^{(2) \, jk}(z) - \frac{g}{2} \varepsilon_{0ijk} H^{(1) \, jk}(z) \right) \dot{z}^i 
\end{align}
and
\begin{align}
m \ddot{z}_i = \bigg( e F_{ij}^{(1)}(z) - g F_{ij}^{(2)}(z) \bigg) \dot{z}^j + \bigg( \frac{e}{2} \varepsilon_{0ijk} H^{(2) \, jk}(z) + \frac{g}{2} \varepsilon_{0ijk} H^{(1) \, jk}(z) \bigg) \dot{z}^0 \, .
\end{align}
Using the solution of the field equation of motion \eqref{eq:m_bh_sol}, the two equations combine into the single equation
\begin{align}
m \ddot{z}_\mu = \left( e F_{\mu\nu}^{(1)}(z) - g F_{\mu\nu}^{(2)}(z) \right) \dot{z}^\nu \, ,
\end{align}
which is equivalent to \eqref{eq:m_eom_z}. All dependencies on the Dirac string drop out since $e \, G_{\mu\nu}^{(2)}(z) - g \, G_{\mu\nu}^{(1)}(z) = 0$. Like the action \eqref{eq:m_sbh_3} this equation is invariant under $\mathrm{SO}(2)$ rotations of the vector potentials $\mathbf{A}^a$ and charges $\mathbf{q}^a$. The variation with respect to $y^0$ and $y^i$ gives only one non-trivial equation
\begin{align}
\begin{aligned}
0 &= e \left( 2 \partial_0 \partial_{[i} A_{j]}^{(1)}(y) - \partial_{[i} \varepsilon_{j]0kl} G^{(2) \, kl}(y) + \partial_{[i} \varepsilon_{j]0kl} H^{(2) \, kl}(y) \right) \\
&\quad - g \left( 2 \partial_0 \partial_{[i} A_{j]}^{(2)}(y) - \partial_{[i} \varepsilon_{j]0kl} G^{(1) \, kl}(y) - \partial_{[i} \varepsilon_{j]0kl} H^{(1) \, kl}(y) \right) \, .
\end{aligned}
\end{align}
We contract this with $\frac{1}{2}\varepsilon^{0ijk}$ and obtain
\begin{align}
0 = e \, \partial_i \left( \varepsilon^{0ijk} \dot{A}_j^{(1)}(y) + G^{(2) \, ik}(y) - H^{(2) \, ik}(y) \right) - g \, \partial_i \left( \varepsilon^{0ijk} \dot{A}_j^{(2)}(y) + G^{(1) \, ik}(y) + H^{(1) \, ik}(y) \right) \, .
\end{align}
Once again $e \, G^{(2) \, ik}(y) - g \, G^{(1) \, ik}(y) = 0$ and so this equation is nothing but the sum of \eqref{eq:m_bh_eom_1} and \eqref{eq:m_bh_eom_2}. In \cite{Deser:1997se} it is a bit vaguely stated that there is no equation for $y_\mu$. A more precise statement would be that the equation of motion for $y_\mu$ is automatically satisfied by any solution to the equations of motion \eqref{eq:m_bh_eom_1} and \eqref{eq:m_bh_eom_2}.

Recalling figure \ref{fig:diag} we have now discussed the arrow from \textbf{Dirac} to \textbf{Bunster-Henneaux}. For the opposed arrow we may start from the action \eqref{eq:m_sbh_3} and reintroduce $A_0^{(1)}$ via the solution to the equation of motion for $A_i^{(2)}$. This will eliminate $A_i^{(2)}$ from the action and yield \eqref{eq:m_sd}.

\subsection{PST}\label{sec:m_pst}
In this subsection, we discuss the PST approach to a manifest duality symmetric action of the Maxwell theory. The original references by Pasti, Sorokin, and Tonin for the content of this section are \cite{Pasti:1995tn, Pasti:1996vs}. The generalization to include sources was first given by Medina and Berkovits in \cite{Medina:1997fn}. Se also \cite{Lechner:1999ga} for a discussion of the PST formalism in the presence of electric and magnetic charge coupling.

We define the generalized twisted self-dual field strengths
\begin{align}\label{eq:m_pst_H}
\mathcal{H}_{\mu\nu}^{(1,2)} &\coloneqq H_{\mu\nu}^{(1,2)} \pm \star H_{\mu\nu}^{(2,1)} \, , \quad \mathcal{H}_{\mu\nu}^{(1,2)} = \pm \star \mathcal{H}_{\mu\nu}^{(2,1)}
\end{align}
and introduce the auxiliary scalar field $a(x)$. It enters the PST action through the non-local timelike harmonic unit vector \eqref{eq:v}
\begin{align}
v_\mu(x) \coloneqq \frac{\partial_\mu a(x)}{\sqrt{- (\partial^\nu a \, \partial_\nu a)}} \, , \quad v_\mu v^\mu = -1 \, ,
\end{align}
with necessarily $\partial_\mu a \not = 0$. There is no systematic derivation of the PST action. Thus, we will just let it fall from the sky and subsequently show that it has all the desired properties. The action reads \cite{Medina:1997fn}
\begin{align}
\begin{aligned}\label{eq:m_spst}
S_{\mathrm{PST}} &= \int \mathrm{d}^4x \ \Big[ -\frac{1}{8} H_{\mu\nu}^{(1)} H^{(1) \, \mu\nu} -\frac{1}{8} H_{\mu\nu}^{(2)} H^{(2) \, \mu\nu} + \frac{1}{4} v^\rho \mathcal{H}_{\rho\mu}^{(1)} \mathcal{H}^{(1) \, \mu\lambda} v_\lambda + \frac{1}{4} v^\rho \mathcal{H}_{\rho\mu}^{(2)} \mathcal{H}^{(2) \, \mu\lambda} v_\lambda \Big] \\
&\quad + \frac{1}{2} I_e - \frac{1}{2} I_g + I_p \, ,
\end{aligned}
\end{align}
with $I_e$ and $I_p$ as above and 
\begin{align}
I_g = g \int \mathrm{d}z^\mu \ A_\mu^{(2)}(z) \, .
\end{align}
Similar to the Bunster-Henneaux action \eqref{eq:m_sbh_3} is invariant under $\mathrm{SO}(2)$ transformations of the vector potentials $A_\mu^{(1,2)}$ and charges. However, it is also manifestly Lorentz invariant. Besides the usual invariance under the gauge transformations $\delta_\mathrm{gauge} A_\mu^{(1,2)} = \partial_\mu \Lambda^{(1,2)}$ the action \eqref{eq:m_spst} has an additional gauge symmetry related to the auxiliary field
\begin{align}\label{eq:m_gauge}
\delta_a A_\mu^{(1,2)} = (\partial_\mu a) \varphi^{(1,2)} \, ,
\quad \delta_a a = 0 \, .
\end{align}
One last time we compute the equations of motion. For $A_\mu^{(1)}$ we obtain
\begin{align}\label{eq:m_pst_eom_1}
0 &= \partial_\nu \left( \mathcal{H}^{(1) \, \mu\nu} - 2 v^{[\mu} \mathcal{H}^{(1) \, \nu]\rho} v_\rho - \varepsilon^{\mu\nu\rho\sigma} v_\rho \mathcal{H}_{\sigma\lambda}^{(2)} v^\lambda \right) \, .
\end{align}
Similarly we find for $A_\mu^{(2)}$
\begin{align}\label{eq:m_pst_eom_2}
0 &= \partial_\nu \left( \mathcal{H}^{(2) \, \mu\nu} - 2 v^{[\mu} \mathcal{H}^{(2) \, \nu]\rho} v_\rho + \varepsilon^{\mu\nu\rho\sigma} v_\rho \mathcal{H}_{\sigma\lambda}^{(1)} v^\lambda \right) \, .
\end{align}
Using $v^{[\mu} \varepsilon^{\nu\rho\sigma\lambda]} = 0$ (which is zero because we anti-symmetrize over five indices in four dimensions) and the twisted self-duality \eqref{eq:m_pst_H} these two equations become
\begin{align}\label{eq:m_pst_eom_3}
0 = \varepsilon^{\mu\nu\rho\sigma} \partial_\nu \left( v_\rho \mathcal{H}_{\sigma\lambda}^{(2,1)} v^\lambda \right) \, .
\end{align}
A general solution of these equations is of the form
\begin{align}
\mathcal{H}_{\mu\nu}^{(1,2)} (\partial^\nu a) = 2 (\partial_{[\mu} \xi^{(1,2)}) (\partial_{\nu]} a) (\partial^\nu a) 
\end{align}
with some functions $\xi^{(1,2)}$. We notice that the right-hand side has the same form as a gauge transformation of the generalized twisted self-dual field strengths, \emph{i.e.}
\begin{align}
\delta_a (\mathcal{H}_{\mu\nu}^{(1,2)} (\partial^\nu a)) = 2 \partial_{[\mu} \varphi^{(1,2)} (\partial_{\nu]} a)(\partial^\nu a) \, .
\end{align}
Thus by fixing the gauge to $\varphi^{(1,2)} = 0$ we obtain $\mathcal{H}_{\mu\nu}^{(1,2)} (\partial^\nu a) = 0$ which is equivalent to 
\begin{align}
\mathcal{H}_{\mu\nu}^{(1,2)} = 0 \, .
\end{align}
This is the same solution to the equations of motion as \eqref{eq:m_bh_sol}. The Maxwell equations are obtained from $\partial_\nu \mathcal{H}^{(1,2) \, \mu\nu} = 0$. It is easy to see that the equation of motion for $A_\mu^{(1)}$ is the Bianchi identity of $H_{\mu\nu}^{(2)}$ and vice versa. The equation of motion for the dyon is as before
\begin{align}
m \ddot{z}_\mu = \left( e F_{\mu\nu}^{(1)}(z) - g F_{\mu\nu}^{(2)}(z) \right) \dot{z}^\nu \, .
\end{align} 
Extremizing \eqref{eq:m_spst} with respect to the string coordinates $y^\mu$ yields the equation
\begin{align}
\begin{aligned}
0 &= e\, \varepsilon^{\mu\nu[\rho\lambda} \partial^{\sigma]} \left( H_{\mu\nu}^{(2)}(y) 
+2 v_\mu \mathcal{H}_{\nu\alpha}^{(2)}(y)v^\alpha + \varepsilon_{\mu\nu\alpha\beta} v^\alpha \mathcal{H}^{(1) \, \beta\tau}(y) v_\tau \right) \\
&\quad + g \, \varepsilon^{\mu\nu[\rho\lambda} \partial^{\sigma]} \left( H_{\mu\nu}^{(1)}(y) 
+2 v_\mu \mathcal{H}_{\nu\alpha}^{(1)}(y)v^\alpha - \varepsilon_{\mu\nu\alpha\beta} v^\alpha \mathcal{H}^{(2) \, \beta\tau}(y) v_\tau \right) \, ,
\end{aligned}
\end{align}
which can be brought into the form
\begin{align}
0 = e \, \varepsilon_{\mu\nu\rho\lambda} \partial^\mu \left( v^\rho \mathcal{H}^{(1) \, \lambda\sigma}(y) v_\sigma \right)
 - g \, \varepsilon_{\mu\nu\rho\lambda} \partial^\mu \left( v^\rho \mathcal{H}^{(2) \, \lambda\sigma}(y) v_\sigma \right) \, .
\end{align}
Again this is the sum of the equations of motion \eqref{eq:m_pst_eom_3}. One can also compute the equation of motion related to the auxiliary field $a(x)$, however, it will also be trivially satisfied by \eqref{eq:m_pst_eom_3}. 

Finally, we briefly discuss the relation of the PST action \eqref{eq:m_spst} to the Dirac and Bunster-Henneaux actions shown in figure \ref{fig:diag}.

\subsubsection*{PST to Dirac}
To obtain the Dirac action from the PST action we plug the solution of the equation of motion for $A_\mu^{(2)}$, \emph{i.e.} $\mathcal{H}_{\mu\nu}^{(1)} (\partial^\nu a) = 0$, into \eqref{eq:m_spst}. In particular, this implies
\begin{align}
\frac{1}{4} v^\rho \mathcal{H}_{\rho\mu}^{(2)} \mathcal{H}^{(2) \, \mu\lambda} v_\lambda = \frac{1}{8} v^\rho v_\rho \mathcal{H}_{\mu\nu}^{(1)} \mathcal{H}^{(1) \, \mu\nu} + \frac{1}{2} v^\rho \mathcal{H}_{\rho\mu}^{(1)} \mathcal{H}^{(1) \, \mu\lambda} v_\lambda 
= - \frac{1}{8} \mathcal{H}_{\mu\nu}^{(1)} \mathcal{H}^{(1) \, \mu\nu} \, .
\end{align}
We then find
\begin{align}
\begin{aligned}
S_\mathrm{PST} &\to \int \mathrm{d}^4x \ \left[ -\frac{1}{8} H_{\mu\nu}^{(1)} H^{(1) \, \mu\nu} -\frac{1}{8} H_{\mu\nu}^{(2)} H^{(2) \, \mu\nu} - \frac{1}{8} \mathcal{H}_{\mu\nu}^{(1)} \mathcal{H}^{(1) \, \mu\nu} \right] 
+ \frac{1}{2} I_e - \frac{1}{2} I_g + I_p \\
&= \int \mathrm{d}^4x \ \left[ -\frac{1}{4} H_{\mu\nu}^{(1)} H^{(1) \, \mu\nu} - \frac{1}{8} \varepsilon^{\mu\nu\rho\lambda} H_{\mu\nu}^{(1)} H_{\rho\lambda}^{(2)} \right] + \frac{1}{2} I_e - \frac{1}{2} I_g + I_p \, .
\end{aligned}
\end{align}
In the second line we use that $\varepsilon^{\mu\nu\rho\lambda} (\star G_{\mu\nu}^{(2)}) (\star G_{\rho\lambda}^{(1)})$ can be regularized to zero \cite{Bekaert:2002eq}. Moreover, notice that $\varepsilon^{\mu\nu\rho\lambda} F_{\mu\nu}^{(1)} F_{\rho\lambda}^{(2)}$ is a total derivative and hence vanishes from the action. Subsequently, we obtain
\begin{align}
\begin{aligned}
S_\mathrm{PST} &\to \int \mathrm{d}^4x \ \left[ -\frac{1}{4} H_{\mu\nu}^{(1)} H^{(1) \, \mu\nu} - \frac{1}{2} A_\nu^{(1)} \partial_\mu G^{(2) \, \mu\nu} - \frac{1}{2} A_\nu^{(2)} \partial_\mu G^{(1) \, \mu\nu} \right] 
+ \frac{1}{2} I_e - \frac{1}{2} I_g + I_p \\
&= - \frac{1}{4} \int \mathrm{d}^4x \ H_{\mu\nu}^{(1)} H^{(1) \, \mu\nu} + I_e + I_p = S_\mathrm{D} \, .
\end{aligned}
\end{align}
\subsubsection*{PST to Bunster-Henneaux}
To reach the Bunster-Henneaux action \eqref{eq:m_sbh_3} we choose $v_\mu = (1,0,0,0)$, which breaks the manifest Lorentz invariance of \eqref{eq:m_spst}. We denote the resulting non Lorentz invariant action by $S_\mathrm{PST}^\mathrm{BH}$ and find
\begin{align}\label{eq:m_spst_bh}
\begin{aligned}
S_\mathrm{PST}^\mathrm{BH}
&= \int \mathrm{d}^4x \ \bigg[ \frac{1}{4} H_{ij}^{(1)} H^{(1) \, ij} + \frac{1}{4} H_{ij}^{(2)} H^{(2) \, ij} 
- \frac{1}{4} \epsilon^{0ijk} H_{0i}^{(1)} H_{jk}^{(2)} + \frac{1}{4} \epsilon^{0ijk} H_{0i}^{(2)} H_{jk}^{(1)} \\
&\quad \quad + \frac{1}{2} F_{0i}^{(2)} G^{(1) \, 0i} - \frac{1}{2} F_{0i}^{(1)} G^{(2) \, 0i} 
+ \frac{1}{4} F_{ij}^{(2)} G^{(1) \, ij} - \frac{1}{4} F_{ij}^{(1)} G^{(2) \, ij} \bigg] + I_p \, .
\end{aligned}
\end{align}
The conjugate momenta are
\begin{align}
\pi^{(1,2) \, i} &= \frac{\partial \mathcal{L}_\mathrm{PST}^\mathrm{BH}}{\partial (\partial_0 A_i^{(1,2)})} = \mp \frac{1}{4} \varepsilon^{0ijk} H_{jk}^{(2,1)} \mp \frac{1}{2} G^{(2,1) \, 0i} = \mp \frac{1}{4} \varepsilon^{0ijk} F_{jk}^{(2,1)} \, ,
\end{align}
where $\mathcal{L}_\mathrm{PST}^\mathrm{BH}$ is the Lagrangian of $S_\mathrm{PST}^\mathrm{BH}$ without $I_p$. The Hamiltonian is
\begin{align}\label{eq:m_pst_bh_h}
\begin{aligned}
\mathcal{H}_\mathrm{PST}^\mathrm{BH} &= F_{0i}^{(1)} \pi^{(1) \, i} + F_{0i}^{(2)} \pi^{(2) \, i} - \mathcal{L}_\mathrm{PST}^\mathrm{BH} \\
&= - \frac{1}{4} H_{ij}^{(1)} H^{(1) \, ij} - \frac{1}{4} H_{ij}^{(2)} H^{(2) \, ij} - \frac{1}{2} F_{ij}^{(2)} G^{(1) \, ij} + \frac{1}{2} F_{ij}^{(1)} G^{(2) \, ij} \, .
\end{aligned}
\end{align}
Subsequently, the canonical action reads
\begin{align}\label{eq:m_pst_can}
S_\mathrm{PST}^\mathrm{can} = \int \mathrm{d}^4x \ \left[ \dot{A}_i^{(1)} \pi^{(1) \, i} + \dot{A}_i^{(2)} \pi^{(2) \, i} - \mathcal{H}_\mathrm{PST}^\mathrm{BH} \right] + I_p \, .
\end{align}
By using the definitions for $\mathbf{B}^a$ \eqref{eq:m_bh_b} and $\boldsymbol{\alpha}^a$ \eqref{eq:m_bh_a} as well as extracting $I_q^\prime$ from the canonical action $S_\mathrm{PST}^\mathrm{can}$ we can transform this expression into \eqref{eq:m_sbh_3}.

All results from this section generalize to $p$-form electrodynamics in $2 (p+1)$ dimensions, \emph{i.e.} those dimensions where dyons can exist. For $p$ odd the equations of motion and Bunster-Henneaux as well as PST actions are invariant under $\mathrm{SO}(2)$ rotations. For $p$ even they are only invariant under $\mathbbm{Z}_2$ transformations of the gauge potentials and charges. For more details see \cite{Deser:1997se,Deser:1997mz,Pasti:1995tn, Pasti:1996vs}. 

\section{Maxwell-Chern-Simons Theory}\label{sec:mcs}
In this section, we discuss electric-magnetic duality in the 5-dimensional Maxwell-Chern-Simons (MCS) theory. We work in flat Minkowski space with mostly plus metric $\eta_{\mu_1\mu_2} = (-, +, \ldots, +)$. The totally anti-symmetric tensor is $\varepsilon_{01234} = 1$. We introduce a 1-form field $A_{\mu_1}^{(1)}$ and its field strength
\begin{align}
F_{\mu_1\mu_2}^{(1)} \coloneqq 2 \partial_{[\mu_1} A_{\mu_2]}^{(1)} \, .
\end{align} 
The starting point of our discussion is the 5-dimensional source-free MCS action
\begin{align}\label{eq:mcs_s}
S = \int \mathrm{d}^5x \ \left[ - \frac{1}{4} F_{\mu_1\mu_2}^{(1)} F^{(1) \, \mu_1\mu_2} + \frac{\alpha}{12} \varepsilon^{\mu_1 \ldots \mu_5} A_{\mu_1}^{(1)} F_{\mu_2\mu_3}^{(1)} F_{\mu_4\mu_5}^{(1)} \right] \, ,
\end{align}
where $\alpha$ is a dimensionful constant. In general MCS theories with cubic Chern-Simons terms exist for $p$-form electrodynamics in $3p + 2$ dimensions with $p$ odd. The equation of motion and Bianchi identity corresponding to the MCS action are
\begin{gather}\label{eq:mcs_eom}
\begin{gathered}
\partial_{\mu_2} F^{(1) \, \mu_1\mu_2} = \frac{\alpha}{4} \varepsilon^{\mu_1 \ldots \mu_5} F_{\mu_2\mu_3}^{(1)} F_{\mu_4\mu_5}^{(1)} \, , \\
\frac{1}{2} \varepsilon^{\mu_1 \ldots \mu_5} \partial_{\mu_3} F_{\mu_4\mu_5}^{(1)} = 0 \, .
\end{gathered}
\end{gather}
These equations give rise to electric-magnetic duality transformations because they can both be written as total derivatives. Moreover, the equations allow for the introduction of electric and magnetic sources. However, compared to the 4-dimensional Maxwell theory the electric and magnetic charges are not carried by a single dyon but rather two separate particles; one electric point charge and one magnetic string charge. The electric point charge (or 0-brane) induces the electric current $J_e^\mu$ and the magnetic string charge (or 1-brane) induces the magnetic current $J_g^{\mu_1\mu_2}$. We will see in a moment that the electric current is in general not conserved in the MCS theory. Thus, we postpone its definition analogous to \eqref{eq:m_J}. The magnetic current, however, is conserved. In analogy to \eqref{eq:m_J} it is parametrized by $z_{(2)} \equiv z_{(2)}(\sigma_1,\sigma_2)$ with $-\infty \le \sigma_{1,2} \le \infty$ and given by
\begin{align}
J_g^{\mu_1\mu_2} \coloneqq g \int \mathrm{d}z_{(2)}^{\mu_1} \wedge \mathrm{d}z_{(2)}^{\mu_2} \ \delta^{(5)}(x-z_{(2)}) \, .
\end{align}
We attach a Dirac 2-brane to the magnetic current which sweeps out a 3-dimensional worldsheet in spacetime. The worldsheet is parametrized by $y_{(2)}(\sigma_1,\sigma_2,\sigma_3)$ with $- \infty \le \sigma_{1,2} \le \infty$ and $0 \le \sigma_3 \le \infty$. We set $y_{(2)}(\sigma_1,\sigma_2,0) \coloneqq z_{(2)}(\sigma_1,\sigma_2)$ such that
\begin{gather}
\star G_{\mu_1\mu_2}^{(2)} = \frac{1}{3!} \varepsilon_{\mu_1 \ldots \mu_5} \, G^{(2) \, \mu_3 \mu_4 \mu_5} \, , \\
G_{\mu_1\mu_2\mu_3}^{(2)}(x) \coloneqq g \int \mathrm{d}y_{(2) \, \mu_1} \wedge \mathrm{d}y_{(2) \, \mu_2} \wedge \mathrm{d}y_{(2) \, \mu_3} \ \delta^{(5)}(x-y_{(2)}) 
\end{gather}
and
\begin{align}
\partial_{\mu_3} G^{(2) \, \mu_1\mu_2\mu_3}(x) = J_g^{\mu_1\mu_2}(x) \, .
\end{align}
Finally, we define the extended field strength
\begin{align}
H_{\mu_1\mu_2}^{(1)} \coloneqq F_{\mu_1\mu_2}^{(1)} + \star G_{\mu_1\mu_2}^{(2)} \, .
\end{align}
Naively adding sources to the equations of motion \eqref{eq:mcs_eom} yields
\begin{gather}\label{eq:mcs_eom_source}
\begin{gathered}
\partial_{\mu_2} H^{(1) \, \mu_1\mu_2} = \frac{\alpha}{4} \varepsilon^{\mu_1 \ldots \mu_5} H_{\mu_2\mu_3}^{(1)} H_{\mu_4\mu_5}^{(1)} + J_e^{\mu_1} \, , \\
\frac{1}{2} \varepsilon^{\mu_1 \ldots \mu_5} \partial_{\mu_3} H_{\mu_4\mu_5}^{(1)} = - J_g^{\mu_1\mu_2} \, .
\end{gathered}
\end{gather}
Taking the exterior derivative of the first equation and using the second one implies
\begin{align}\label{eq:mcs_eom_ext_d}
\partial_{\mu_1} J_e^{\mu_1} = - \alpha H_{\mu_1\mu_2}^{(1)} J_g^{\mu_1\mu_2} \, .
\end{align}
Thus, the electric current is not conserved on the worldsheet of the magnetic string. Moreover, the conservation rule for $J_e^{\mu_1}$ explicitly depends on the field strength $H_{\mu_1\mu_2}^{(1)}$. This prohibits us from specifying the electric current and severely complicates the formulation of an action principle giving rise to \eqref{eq:mcs_eom_source} \cite{Bekaert:2002eq}. To circumvent the difficulties one can either directly introduce boundary conditions on the worldsheet of the magnetic string, or define an additional field living on the worldsheet, which will then dynamically introduce the boundary conditions. In the following, we choose the latter approach as in \cite{Bekaert:2002eq}.

\subsubsection*{The Field on the Brane}
Let $w^{M_1}$ be a set of coordinates defined on the 2-dimensional worldsheet $\mathcal{M}_2$ of the magnetic string. The metric is $\eta_{M_1M_2} = (-, +)$ and the antisymmetric symbol is $\epsilon_{01} = 1$. Moreover, let
\begin{align}
f_{M_1}(w) \coloneqq \partial_{M_1} b(w)- A_{M_1}^{(1) \, \ast}(w) 
\end{align}
be the field strength of the worldsheet gauge field $b(w)$. $A_{M_1}^{(1) \, \ast}(w)$ denotes the pullback of $A_{\mu_1}^{(1)}(x)$ on the worldsheet $\mathcal{M}_2$. The field strength $f_{M_1}$ is invariant under the gauge transformations
\begin{align}\label{eq:mcs_f_gauge}
\delta_\mathrm{gauge} \, A_{\mu_1}^{(1)} = \partial_{\mu_1} \Lambda^{(1)} \, , \quad \delta_\mathrm{gauge} \, b = \Lambda^{(1) \, \ast} \, .
\end{align}
If the worldline of the electric point charge ends on the magnetic string it produces an instanton, \emph{i.e.} a (-1)-brane. We denote the dyonic instanton current by $j(w)$. This current does not sweep out any kind of worldsheet but only exists at the fixed spacetime point $z_{(3)}$, where the electric point charge hits the magnetic string. From this spacetime point originates a Dirac 0-brane parametrized by $y_{(3)} \equiv y_{(3)}(\sigma)$ with $0 \le \sigma \le \infty$ and $y_{(3)}(0) \coloneqq z_{(3)}$ living on the worldsheet of the magnetic string (see figure \ref{fig:instanton}).\footnote{In 11-dimension the 2-brane can end on the 5-brane and produce a string charge which then lives together with its attached Dirac 2-brane on the worldsheet of the 5-brane.} The Dirac 1-brane is defined on $\mathcal{M}_2$ as
\begin{gather}
\star_2 \, g_{M_1} = \epsilon_{M_1M_2} g^{M_2} \, , \\
g^{M_1} \coloneqq q \int \mathrm{d}y_{(3)}^{M_1} \ \delta^{(2)}(w-y_{(3)}) \, ,
\end{gather}
where $\star_2$ denotes the Hodge dual on $\mathcal{M}_2$ and $q$ is the dyonic instanton charge. As usual, the Dirac 0-brane satisfies the equation
\begin{align}\label{eq:mcs_j}
\partial_{M_1} g^{M_1}(w) = q \, \delta^{(2)}(w-y_{(3)}) \eqqcolon j(w) \, . 
\end{align}
The extended field strength on the worldvolume is
\begin{align}
h_{M_1}(w) \coloneqq f_{M_1}(w) + \star_2 \, g_{M_1}(w) \, . 
\end{align}
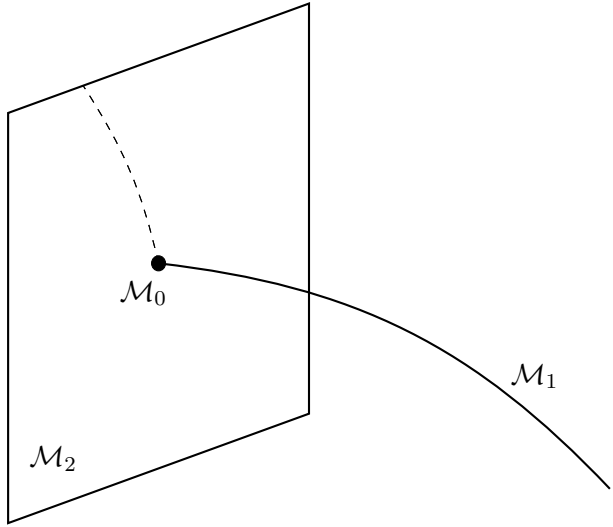
\begin{figure}[t]
\begin{tikzpicture}
\node [trapezium, trapezium left angle=-70, trapezium right angle=-110, rotate=90, minimum width=4cm, minimum height=4cm, draw=black, thick] at (0,0) { };
\draw [bend left=20, thick] (0,0) to (6,-3);
\draw [bend right=10, dashed] (0,0) to (-1,2.36);
\draw [bend right=10, dashed] (0,0) to (-1,2.36);
\draw [fill=black] (0,0) ellipse (0.093 and 0.1);
\node at (-.2,-.4) {$\mathcal{M}_0$};
\node at (5,-1.5) {$\mathcal{M}_1$};
\node at (-1.4,-2.6) {$\mathcal{M}_2$};
\end{tikzpicture}
\caption{$\mathcal{M}_2$ is the worldsheet of the magnetic string. The worldline $\mathcal{M}_1$ of the electric point charge ended on the magnetic string at some spacetime point $\mathcal{M}_0$ and produced an instanton (black dot). From $\mathcal{M}_0$ emerges the worldline of a Dirac point (dashed line), which lives on the worldsheet of the magnetic string $\mathcal{M}_2$ \cite{Bekaert:2002eq}.}
\label{fig:instanton}
\end{figure}

\subsection{Dirac}
The Dirac action of the 5-dimensional MCS theory coupled to an electric point charge and a magnetic string charge is \cite{Bekaert:2002eq}
\begin{align}\label{eq:mcs_sd}
\begin{aligned}
S_D &= \int \mathrm{d}^5 x \ \left[ - \frac{1}{4} H_{\mu_1\mu_2}^{(1)} H^{(1) \, \mu_1\mu_2} + \frac{\alpha}{12}\varepsilon^{\mu_1 \ldots \mu_5} A_{\mu_1}^{(1)} F_{\mu_2\mu_3}^{(1)} F_{\mu_4\mu_5}^{(1)} + \frac{\alpha}{4} \varepsilon^{\mu_1 \ldots \mu_5} A_{\mu_1}^{(1)} F_{\mu_2\mu_3}^{(1)} (\star G_{\mu_4\mu_5}^{(2)}) \right] \\
&\quad - \alpha g \int \mathrm{d}w^{M_1} \wedge \mathrm{d}w^{M_2} \, \left[ - h_{M_1} A_{M_2}^{(1) \, \ast} + \frac{1}{2} (\star_2 \, h_{M_1}) h_{M_2} + \frac{\beta}{2} \, (\star_2 \, j_{M_1M_2}) b \right] + I_e + I_p \, ,
\end{aligned}
\end{align}
with a dimensionless constant $\beta$ and
\begin{gather}
I_e = \int \mathrm{d}^5x \ A_{\mu_1}^{(1)} J_e^{\mu_1} \, , 
\label{eq:mcs_d_Ie} \\
I_p = \sum_{k=1}^2 m_k \int \mathrm{d}^k\sigma \ \sqrt{-\gamma_{(k)}} \left( \gamma_{(k)}^{IJ} \partial_I z_{(k)}^{\mu_1} \partial_J z_{(k) \, \mu_1} - (k-1) \right) \, .
\label{eq:mcs_d_Ip}
\end{gather}
Because the sources are 0- and 1-branes we have replaced the action of a free point particle \eqref{eq:m_d_Ip} by the more general Polyakov type action \eqref{eq:mcs_d_Ip}. There are two terms, one for each brane. $\gamma_{(k)}^{IJ}$ are the metrics on the worldvolumes of the electric and magnetic branes. The indices $I$, $J$ run from $0$ to $k-1$. The electric charge has mass $m_1$ and the magnetic charge has mass $m_2$.

We can extend the integral over the magnetic worldsheet in \eqref{eq:mcs_sd} to an integral over the 5-dimensional spacetime by inserting a closed 2-form derived from the magnetic current \cite{Bandos:1997gd}, such that
\begin{gather}
\begin{gathered}
 \int \mathrm{d}w^{M_1} \wedge \mathrm{d}w^{M_2} \, \left[ - h_{M_1} A_{M_2}^{(1) \, \ast} + \frac{1}{2} (\star_2 \, h_{M_1}) h_{M_2} + \frac{\beta}{2} \, (\star_2 \, j_{M_1M_2}) b \right] \\
 = \int \mathrm{d}^5x \ \left( - h_{\mu_1} A_{\mu_2}^{(1)} + \frac{1}{2} (\star_2 \, h_{\mu_1}) h_{\mu_2} + \frac{\beta}{2} \, (\star_2 \, j_{\mu_1\mu_2}) b \right) J_g^{\mu_1\mu_2} \, .
\end{gathered}
\end{gather}
This term has been added to the action \eqref{eq:mcs_sd} to modify the naive equations of motion \eqref{eq:mcs_eom_source} in such a way that the right-hand side of \eqref{eq:mcs_eom_ext_d} no longer depends on the field strength $H_{\mu_1\mu_2}^{(1)}$. The instanton current has been added to account for the situation described in figure \ref{fig:instanton}. Demanding that the action \eqref{eq:mcs_sd} is invariant under the gauge transformations \eqref{eq:mcs_f_gauge} yields the equation
\begin{align}\label{eq:mcs_d_gauge}
\partial_{\mu_1} J_e^{\mu_1} = \frac{\alpha(1- \beta)}{2} (\star_2 \, j_{\mu_1\mu_2}) J_g^{\mu_1\mu_2} \, .
\end{align}
Thus for $\beta \not= 1$, the electric current is not conserved if the electric worldline ends on the magnetic string and produces a dyonic instanton. The equation of motion for the gauge field $A_{\mu_1}^{(1)}$ is easily computed from the action \eqref{eq:mcs_sd} and reads
\begin{align}\label{eq:mcs_d_eom}
\partial_{\mu_2} H^{(1) \, \mu_1 \mu_2} = \frac{\alpha}{4} \varepsilon^{\mu_1 \ldots \mu_5} H_{\mu_2\mu_3}^{(1)} H_{\mu_4 \mu_5}^{(1)} - \alpha (h_{\mu_2} + (\star_2 \, h_{\mu_2})) J_g^{\mu_1\mu_2} + J_e^{\mu_1} \, .
\end{align}
We have added a term $\varepsilon^{\mu_1 \ldots \mu_5}(\star G_{\mu_2\mu_3}^{(2)}) (\star G_{\mu_4\mu_5}^{(2)})$ which can be regularized away \cite{Bekaert:2002eq}. The corresponding Bianchi identity is
\begin{align}\label{eq:mcs_d_bianchi}
\frac{1}{2}\varepsilon^{\mu_1 \ldots \mu_5} \partial_{\mu_3} H_{\mu_4\mu_5}^{(1)} = - J_g^{\mu_1\mu_2} \, .
\end{align}
The equation of motion for the auxiliary field $b$ and the Bianchi identity of its field strength are
\begin{align}\label{eq:mcs_d_eom_b}
\begin{aligned}
\partial_{M_1} h^{M_1} &= - \frac{1}{2} \epsilon^{M_1M_2} F_{M_1M_2}^{(1) \, \ast} - \frac{\beta}{2} \, j \, , \\
\partial_{M_1} (\star_2 \, h^{M_1}) &= - \frac{1}{2} \epsilon^{M_1M_2} F_{M_1M_2}^{(1) \, \ast} + \frac{1}{2} \, j \, .
\end{aligned}
\end{align}
As a consistency check, we take the exterior derivative of the equation of motion \eqref{eq:mcs_d_eom} and find using \eqref{eq:mcs_d_eom_b}
\begin{align}
0 = \partial_{\mu_1} \partial_{\mu_2} H^{(1) \, \mu_1 \mu_2} = - \alpha F_{\mu_1\mu_2}^{(1)} J_g^{\mu_1\mu_2} + \alpha F_{\mu_1\mu_2}^{(1)} J_g^{\mu_1\mu_2} - \frac{\alpha(1- \beta)}{2} (\star_2 \, j_{\mu_1\mu_2}) J_g^{\mu_1\mu_2} + \partial_{\mu_1} J_e^{\mu_1} \, .
\end{align}
Again we have assumed that $\varepsilon^{\mu_1 \ldots \mu_5}(\star G_{\mu_2\mu_3}^{(2)}) (\star G_{\mu_4\mu_5}^{(2)}) = 0$. This equation is equivalent to the gauge invariance condition \eqref{eq:mcs_d_gauge}. Hence the action \eqref{eq:mcs_sd} is self-consistent and the electric current is well-defined.

Because the worldsheet of the magnetic string is a Minkowski manifold we can decompose the field strength $h_{M_1}$ into its self-dual part $h_{M_1}^+$ and anti-self dual part $h_{M_1}^-$
\begin{align}
h_{M_1} = h_{M_1}^+ + h_{M_1}^- \, , \quad \star_2 \, h_{M_1}^\pm = \pm h_{M_1}^\pm \, .
\end{align}
The equations \eqref{eq:mcs_d_eom_b} are then equivalent to 
\begin{align}\label{eq:mcs_d_eom_h+-}
\begin{aligned}
\partial_{M_1} h^{+ \, M_1} &= - \frac{1}{2}\epsilon^{M_1M_2} F_{M_1M_2}^{(1) \, \ast} + \frac{1-\beta}{2} \, j \, , \\
\partial_{M_1} h^{- \, M_1} &= - \frac{1+\beta}{2} \, j \, .
\end{aligned}
\end{align}
For $\beta =-1$ it is natural to consider the case where $h_{M_1}^- = 0$, \emph{i.e.} in which the field strength is self-dual. We will see later that this is the necessary case when building up the PST or Bunster-Henneaux versions of the action \eqref{eq:mcs_sd}. Moreover, this is also exactly the case that appears in supergravity. 

Besides the field equations of motion \eqref{eq:mcs_d_eom} - \eqref{eq:mcs_d_eom_b} there are also equations of motion derived from the variation with respect to the string variables $y_{(2,3)}^{\mu_1}$ and $z_{(1,2)}^{\mu_1}$. There is no variation with respect to $z_{(3)}^{\mu_1}$ because the instanton does not even trace out a worldline. Computing the variations as shown in \cite{Dirac:1948um} we find the following equation for $z_{(1)}^{\mu_1}$
\begin{align}\label{eq:mcs_d_eom_z1}
m_1 \, \partial_J \left( \sqrt{-\gamma_{(1)}} \gamma_{(1)}^{IJ} \partial_I z_{(1) \, \mu_{1}} \right) = \frac{e}{2} F_{\mu_1\mu_2}^{(1)}(z_{(1)}) \frac{\partial z_{(1)}^{\mu_2}}{\partial \sigma_1} \, .
\end{align}
This is the analog of \eqref{eq:m_d_eom_z} for an electric point particle moving in an electromagnetic field. For $j = 0$ the equation of motion for $z_{(2)}^{\mu_1}$ reads\footnote{The choice $j=0$ will be justified by the considerations in the next subsection.}
\begin{align}
\begin{aligned}\label{eq:mcs_d_eom_z2}
m_2 \, \partial_J &\left( \sqrt{-\gamma_{(2)}} \gamma_{(2)}^{IJ} \partial_I z_{(2)}^{\mu_{1}} \right) \\
&= \frac{g}{4} \varepsilon^{\mu_1 \ldots \mu_5} \bigg( H_{\mu_4\mu_5}^{(1)}(z_{(2)}) + \alpha \, \varepsilon_{\mu_4 \ldots \mu_8} h^{+ \, \mu_6}(z_{(2)}) F^{(1) \, \mu_7\mu_8}(z_{(2)}) \bigg) \frac{\partial z_{(2) \, \mu_2}}{\partial \sigma_1} \frac{\partial z_{(2) \, \mu_3}}{\partial \sigma_2} \, .
\end{aligned}
\end{align}
The second term comes from the auxiliary field term in \eqref{eq:mcs_sd}. The equation of motion for $ y_{(2)}^{\mu_1}$ is
\begin{align}\label{eq:mcs_d_eom_y2}
0 = g \, \partial_{\mu_2} \left( H^{(1) \, \mu_1\mu_2}(y_{(2)}) - \frac{\alpha}{2} \varepsilon^{\mu_1 \ldots \mu_5} A_{\mu_3}^{(1)}(y_{(2)}) F_{\mu_4\mu_5}^{(1)}(y_{(2)}) \right) \, .
\end{align}
This is equivalent to \eqref{eq:m_d_eom_y} if we assume that $J_e^{\mu_1}(y_{(2)}) = 0$, $J_g^{\mu_1\mu_2}(y_{(2)}) = 0$ as well as $\star G_{\mu_1\mu_2}^{(2)}(y_{(2)}) = 0$. These three conditions yield the three Dirac vetos stated below. Finally, the equation of motion for $y_{(3)}^{M_1}$ is
\begin{align}
\partial_{M_1} h^{M_1}(y_{(3)}) = - \frac{1}{2} \epsilon^{M_1M_2} F_{M_1M_2}^{(1) \, \ast}(y_{(3)}) \, .
\end{align}
This is the same as the first equation in \eqref{eq:mcs_d_eom_b} for $j(y_{(3)}) = 0$. 

\subsubsection*{Dirac Vetos}
To summarize the string equations imply the following Dirac vetos \cite{Bekaert:2002eq}:
\begin{itemize}
\item No electric charge can be located on the Dirac worldvolume. This is the standard Dirac veto.
\item No magnetic charge can be located on the Dirac worldvolume. 
\item No dyonic instanton charge can be located on the Dirac worldline.
\item Dirac worldvolumes cannot intersect each other (or themselves). In other words, the pullback of $\star G_{\mu_1\mu_2}^{(2)}$ on the Dirac wolrdvolume must vanish. 
\end{itemize}
Note that the Dirac vetos, in the presence of the Chern-Simons term, are more restrictive than in the Maxwell case where only the first one applies. 

\subsection{PST}\label{sec:mcs_pst}
We discuss the PST approach to a duality symmetric formulation of the 5-dimensional MCS theory. Because the MCS theory is considerably more complicated than the pure Maxwell theory, we skip the derivation of the Bunster-Henneaux action from the Dirac action \eqref{eq:mcs_sd}. Instead, we will make use of figure \ref{fig:diag} and derive it from the PST action.

In theories with a Chern-Simons term, the electric-magnetic duality relation is no longer between the gauge potentials but between their field strengths. This implies that it cannot be made manifest in the action. To introduce the dual gauge potential and its field strength we first consider the MCS theory without sources \eqref{eq:mcs_s}. The equation of motion for $A_{\mu_1}^{(1)}$ \eqref{eq:mcs_eom} must imply the Bianchi identity of the dual field strength $F_{\mu_1\mu_2\mu_3}^{(2)}$, \emph{i.e.}
\begin{align}\label{eq:mcs_pst_bianchi_F2}
- \frac{1}{3!} \varepsilon^{\mu_1 \ldots \mu_5} \partial_{\mu_2} F_{\mu_3\mu_4\mu_5}^{(2)} = \frac{\alpha}{4} \varepsilon^{\mu_1 \ldots \mu_5} F_{\mu_2\mu_3}^{(1)} F_{\mu_4\mu_5}^{(1)} \, .
\end{align}
We can give a general solution for $F_{\mu_1\mu_2\mu_3}^{(2)}$ in terms of a dual 2-form field $A_{\mu_1\mu_2}^{(2)}$ and the 1-form field $A_{\mu_1}^{(1)}$
\begin{align}
F_{\mu_1\mu_2\mu_3}^{(2)} \coloneqq 3 \partial_{[\mu_1} A_{\mu_2\mu_3]}^{(2)} - 3 \alpha A_{[\mu_1}^{(1)} F_{\mu_2\mu_3]}^{(1)} \, .
\end{align}
To find the extended dual field strength $H_{\mu_1\mu_2\mu_3}^{(2)}$ we consider \eqref{eq:mcs_d_eom} instead of \eqref{eq:mcs_eom} such that \eqref{eq:mcs_pst_bianchi_F2} becomes
\begin{align}\label{eq:mcs_pst_bianchi_H2}
- \frac{1}{3!} \varepsilon^{\mu_1 \ldots \mu_5} \partial_{\mu_2} H_{\mu_3\mu_4\mu_5}^{(2)} = \frac{\alpha}{4} \varepsilon^{\mu_1 \ldots \mu_5} H_{\mu_2\mu_3}^{(1)} H_{\mu_4 \mu_5}^{(1)} - 2 \alpha h_{\mu_2}^+ J_g^{\mu_1\mu_2} + J_e^{\mu_1} \, .
\end{align}
This equation has a solution in terms of the dual field strength $F_{\mu_1\mu_2\mu_3}^{(2)}$, the auxiliary field strength $h_{\mu_1}$, the Dirac 2-brane $\star G_{\mu_1\mu_2}^{(2)}$ and a Dirac 1-brane $\star G_{\mu_1\mu_2\mu_3}^{(1)}$. Usually $\star G_{\mu_1\mu_2\mu_3}^{(1)}$ would be the Dirac string attached to the electric current. However, since the electric current is in general not conserved (see \eqref{eq:mcs_d_gauge}) the Dirac string $\star G_{\mu_1\mu_2\mu_3}^{(1)}$ must be attached to some more general object than the electric point charge alone. We will derive its properties from the solution of \eqref{eq:mcs_pst_bianchi_H2}. We make the general ansatz
\begin{align}
H_{\mu_1\mu_2\mu_3}^{(2)} \coloneqq 3 \partial_{[\mu_1} A_{\mu_2\mu_3]}^{(2)} + \star G_{\mu_1\mu_2\mu_3}^{(1)} - 3 \alpha A_{[\mu_1}^{(1)} F_{\mu_2\mu_3]}^{(1)} + 6 \alpha h_{[\mu_1} (\star G_{\mu_2\mu_3]}^{(2)})
\end{align}
and compute
\begin{align}
\begin{aligned}\label{eq:mcs_pst_bianchi_H2_sol}
- \frac{1}{3!} \varepsilon^{\mu_1 \ldots \mu_5} \partial_{\mu_2} H_{\mu_3\mu_4\mu_5}^{(2)} &= - \frac{1}{2} \varepsilon^{\mu_1 \ldots \mu_5} \partial_{\mu_2} \partial_{\mu_3} A_{\mu_4\mu_5}^{(2)} - \frac{1}{3!} \varepsilon^{\mu_1 \ldots \mu_5} \partial_{\mu_2} (\star G_{\mu_3\mu_4\mu_5}^{(1)}) \\
&\quad + \frac{\alpha}{2} \varepsilon^{\mu_1 \ldots \mu_5} \partial_{\mu_2} \left( A_{\mu_3}^{(1)} F_{\mu_4\mu_5}^{(1)} \right) - \alpha \varepsilon^{\mu_1 \ldots \mu_5} \partial_{\mu_2} \left( h_{\mu_3} (\star G_{\mu_4\mu_5}^{(2)}) \right) \\
&= - \frac{1}{3!} \varepsilon^{\mu_1 \ldots \mu_5} \partial_{\mu_2} (\star G_{\mu_3\mu_4\mu_5}^{(1)}) + \frac{\alpha}{4} \varepsilon^{\mu_1 \ldots \mu_5} H_{\mu_2\mu_3}^{(1)} H_{\mu_4 \mu_5}^{(1)} \\
&\quad - \frac{\alpha}{2} \varepsilon^{\mu_1 \ldots \mu_5} (\star_2 \, j_{\mu_2\mu_3}) (\star G_{\mu_4\mu_5}^{(2)}) - 2 \alpha h_{\mu_2} J_g^{\mu_1\mu_2} \, .
\end{aligned}
\end{align}
Here we have used the Bianchi identity for $h_{\mu_1}$ \eqref{eq:mcs_d_eom_b}. Comparing the right hand sides of \eqref{eq:mcs_pst_bianchi_H2} and \eqref{eq:mcs_pst_bianchi_H2_sol} we first observe that $h_{\mu_1}$ must be self-dual, \emph{i.e.} $h_{\mu_1} = h_{\mu_1}^+$. By \eqref{eq:mcs_d_eom_h+-} this implies $\beta = -1$. For a consistent PST formulation, it is crucial that the self-duality condition $h_{\mu_1} = h_{\mu_1}^+$ arises from the equations of motion. From comparing \eqref{eq:mcs_pst_bianchi_H2} and \eqref{eq:mcs_pst_bianchi_H2_sol} we then also find
\begin{align}
\partial_{\mu_2} G^{(1) \, \mu_1\mu_2} = J_e^{\mu_1} + \frac{\alpha}{2} \varepsilon^{\mu_1 \ldots \mu_5} (\star_2 \, j_{\mu_2\mu_3}) (\star G_{\mu_4\mu_5}^{(2)}) \, .
\end{align}
The exterior derivative of this equation gives
\begin{align}
0 = \partial_{\mu_1} \partial_{\mu_2} G^{(1) \, \mu_1\mu_2} = \partial_{\mu_1} J_e^{\mu_1} - \alpha (\star_2 \, j_{\mu_1\mu_2}) J_g^{\mu_1\mu_2} + \frac{\alpha}{2} \varepsilon^{\mu_1 \ldots \mu_5} \partial_{\mu_1} (\star_2 \, j_{\mu_2\mu_3}) (\star G_{\mu_4\mu_5}^{(2)}) \, .
\end{align}
The first two terms on the right-hand side are the gauge condition \eqref{eq:mcs_d_gauge} with $\beta = -1$. The third term vanishes if the instanton current is conserved. By \eqref{eq:mcs_j} this is only the case for $j = 0$. In 11 dimensions the current on the M5-brane is a 1-brane and thus automatically conserved. However, in the 5-dimensional theory, we are (for the PST and Bunster-Henneaux formulations) limited to the case where the electric and magnetic branes do not intersect. Thus we set $j = 0$ and $g_{M_1} = 0$ and subsequently replace $h_{M_1}$ with $f_{M_1}$. This choice implies that the electric current $J_e^{\mu_1}$ is conserved. The extended dual field strength is then given by
\begin{align}
H_{\mu_1\mu_2\mu_3}^{(2)} \coloneqq 3 \partial_{[\mu_1} A_{\mu_2\mu_3]}^{(2)} + \star G_{\mu_1\mu_2\mu_3}^{(1)} - 3 \alpha A_{[\mu_1}^{(1)} F_{\mu_2\mu_3]}^{(1)} + 6 \alpha f_{[\mu_1} (\star G_{\mu_2\mu_3]}^{(2)}) \, , 
\end{align}
where now $\star G_{\mu_1\mu_2\mu_3}^{(1)}$ is the Dirac string attached to the electric point charge only. We parametrize the worldline of the electric current by $z_{(1)}(\sigma_1)$ with $- \infty \le \sigma_1 \le \infty$. The Dirac string is parametrized by $y_{(2)}(\sigma_1, \sigma_2)$ with $- \infty \le \sigma_1 \le \infty$, $0 \le \sigma_2 \le \infty$ and $y_{(2)}(\sigma_1, 0) \coloneqq z_{(2)}(\sigma_1)$. We have
\begin{gather}
\begin{gathered}
\star G_{\mu_1\mu_2\mu_3}^{(1)} = \frac{1}{2} \varepsilon_{\mu_1 \ldots \mu_5} G^{(1) \, \mu_4\mu_5} \, , \\
G_{\mu_1\mu_2}^{(1)} (x) \coloneqq e \int \mathrm{d}y_{(1) \, \mu_1} \wedge \mathrm{d}y_{(1) \, \mu_2} \ \delta^{(5)}(x-y_{(1)})
\end{gathered}
\end{gather}
and
\begin{align}
\partial_{\mu_2} G^{(1) \mu_1\mu_2}(x) = J_e^{\mu_1}(x) \, .
\end{align}
Finally, we introduce the duality-related generalized field strengths
\begin{align}
\mathcal{H}_{\mu_1\mu_2}^{(1)} \coloneqq H_{\mu_1\mu_2}^{(1)} + \star H_{\mu_1\mu_2}^{(2)} \, , \quad \quad
\mathcal{H}_{\mu_1\mu_2\mu_3}^{(2)} \coloneqq H_{\mu_1\mu_2\mu_3}^{(2)} - \star H_{\mu_1\mu_2\mu_3}^{(1)} 
\end{align}
with
\begin{align}
\mathcal{H}_{\mu_1\mu_2}^{(1)} = - \star \mathcal{H}_{\mu_1\mu_2}^{(2)} \, , \quad \quad 
\mathcal{H}_{\mu_1\mu_2\mu_3}^{(2)} = \star \mathcal{H}_{\mu_1\mu_2\mu_3}^{(2)} \, .
\end{align}
When the duality relation $H_{\mu_1\mu_2}^{(1)} = - \star H_{\mu_1\mu_2}^{(2)}$ is satisfied the generalized field strengths vanish. The PST action corresponding to \eqref{eq:mcs_sd} then reads
\begin{align}
\begin{aligned}\label{eq:mcs_spst}
S_\mathrm{PST} &= \int \mathrm{d}^5 x \ \bigg[ - \frac{1}{8} H_{\mu_1\mu_2}^{(1)} H^{(1) \, \mu_1\mu_2} - \frac{1}{24} H_{\mu_1\mu_2\mu_3}^{(2)} H^{(2) \, \mu_1\mu_2\mu_3} \\
&\quad + \frac{1}{4} v^\rho \mathcal{H}_{\rho\mu_1}^{(1)} \mathcal{H}^{(1) \, \mu_1\lambda} v_\lambda - \frac{1}{8} v^\rho \mathcal{H}_{\rho\mu_1\mu_2}^{(2)} \mathcal{H}^{(2) \, \mu_1\mu_2\lambda} v_\lambda \\
&\quad + \frac{1}{72} \varepsilon^{\mu_1 \ldots \mu_5} F_{\mu_1\mu_2\mu_3}^{(2)} F_{\mu_4\mu_5}^{(1)} 
+ \frac{\alpha}{8} \varepsilon^{\mu_1 \ldots \mu_5} f_{\mu_1} F_{\mu_2\mu_3}^{(1)} (\star G_{\mu_4\mu_5}^{(2)}) \bigg] 
+ \frac{1}{2} I_e + \frac{1}{2} I_g + I_p \, , 
\end{aligned}
\end{align}
with $I_e$ and $I_p$ defined in \eqref{eq:mcs_d_Ie} - \eqref{eq:mcs_d_Ip} and
\begin{align}\label{eq:mcs_pst_Ig}
\begin{aligned}
I_g &= \alpha \int \mathrm{d}^5x \ \left( \frac{1}{2\alpha} A_{\mu_1\mu_2}^{(2)} + f_{\mu_1} A_{\mu_2}^{(1)} \right) J_g^{\mu_1\mu_2} \\
&\quad +\alpha \int \mathrm{d}^2 w \ \left[ f_{M_1} f^{M_1} - v^{M_1} (f_{M_1} - (\star_2 \, f_{M_1})) (f^{M_2} - (\star_2 \, f^{M_2})) v_{M_2} \right] \, .
\end{aligned}
\end{align}
The action \eqref{eq:mcs_spst} is the first main result of this paper. The first two lines of the action are as expected when compared to the corresponding action for the Maxwell theory \eqref{eq:m_spst}. In the Chern-Simons terms we replaced $A_{\mu_1}^{(1)} F_{\mu_2\mu_3}^{(1)}$ with $F_{\mu_1\mu_2\mu_3}^{(2)}$ and $f_{\mu_1} F_{\mu_2\mu_3}^{(1)}$ respectively. Moreover, the coefficients in front of these terms are different to the ones of the respective terms in \eqref{eq:mcs_sd}. This is because also a part of the Chern-Simons terms is now hidden in the generalized field strength terms in the second line of \eqref{eq:mcs_spst}.

The action $I_g$ describes the coupling of the magnetic string to both the gauge potential $A_{\mu_1}^{(1)}$ and its dual $A_{\mu_1\mu_2}^{(2)}$. It couples to both fields because $J_g^{\mu_1\mu_2}$ appears in both \eqref{eq:mcs_d_eom} and \eqref{eq:mcs_d_bianchi}. A part of this action was already present in \eqref{eq:mcs_sd}. The last term in \eqref{eq:mcs_pst_Ig} has been added to ensure the self-duality of the auxiliary field strength via the equation of motion for $b$. Remarkably $I_g$ agrees exactly with the linearized 5-brane action in supergravity \cite{Pasti:1997gx}. 

All these changes are such that the equations of motion of \eqref{eq:mcs_spst} are solved by the duality relations of $H_{\mu_1\mu_2}^{(1)}$, $H_{\mu_1\mu_2\mu_3}^{(2)}$ and $f_{M_1}$, as shown below.

Compared to the corresponding expression for 11-dimensional supergravity coupled to the M2- and M5-brane in \cite{Bandos:1997gd} we have a different sign in front of the term
\begin{align}\label{eq:mcs_pst_aux1}
\frac{\alpha}{8} \varepsilon^{\mu_1 \ldots \mu_5} f_{\mu_1} F_{\mu_2\mu_3}^{(1)} (\star G_{\mu_4\mu_5}^{(2)}) \, .
\end{align}
This difference is crucial for the consistency of the action. The authors of \cite{Bandos:1997gd} emphasize that their version of the PST action yields consistent equations of motion without assuming $(\star G^{(2)}) \wedge (\star G^{(2)}) = 0$ as we have done here. This is true for the PST action alone. However, when the PST action of \cite{Bandos:1997gd} is reduced to its corresponding Dirac action (by the means of figure \ref{fig:diag}) a contradiction between the field and string equations of motion arises. In the corresponding equations to \eqref{eq:mcs_d_eom} and \eqref{eq:mcs_d_eom_y2} derived from the action given at the end of chapter 5 in \cite{Bandos:1997gd} the coefficients in front of the Chern-Simons terms do not match. Thus the magnetic brane becomes dynamical and the only solution consistent with both equations in \cite{Bandos:1997gd} is $F^{(1)} = 0$. If our sign for \eqref{eq:mcs_pst_aux1} is adopted and one assumes that $(\star G^{(2)}) \wedge (\star G^{(2)}) = 0$ the contradiction in \cite{Bandos:1997gd} disappears. The reader should take this as a warning when working with Dirac branes to be very careful in computing all the equations of motion and in particular check that the string equations of motion are consistent with the field equations of motion. 

Finally, let us mention that contrary to the claim made in \cite{Bandos:1997gd} the action (even without sources) is not invariant under the simultaneous exchange of
\begin{align}
F_{\mu_1\mu_2}^{(1)} \to - \frac{1}{3!} \varepsilon_{\mu_1 \ldots \mu_5} F^{(2) \, \mu_3\mu_4\mu_5} \, , \quad F_{\mu_1\mu_2\mu_3}^{(2)} \to \frac{1}{2} \varepsilon_{\mu_1 \ldots \mu_5} F^{(1) \, \mu_4\mu_5} \, .
\end{align}
While the terms in the first line of \eqref{eq:mcs_spst} pick up a minus sign the terms containing the generalized field strengths and the Chern-Simons term are invariant under this transformation. Hence twisted self-duality for the MCS theory is something that can only exist at the level of the equations of motion.

The action \eqref{eq:mcs_spst} is invariant under the usual gauge transformations
\begin{gather}
\begin{gathered}
\delta_\mathrm{gauge} \, A_{\mu_1}^{(1)} = \partial_{\mu_1} \Lambda^{(1)} \, , 
\quad
\delta_\mathrm{gauge} \, A_{\mu_1\mu_2}^{(2)} = 2 \partial_{[\mu_1} \Lambda_{\mu_2]}^{(2)} + 2 \alpha \partial_{[\mu_1} A_{\mu_2]}^{(1)} \Lambda^{(1)} \, , 
\\
\delta_\mathrm{gauge} \, b = \Lambda^{(1)} \, , 
\quad
\delta_\mathrm{gauge} \, a = 0 \, .
\end{gathered}
\end{gather}
Moreover the action \eqref{eq:mcs_spst} is invariant under the special gauge transformations\footnote{In the case of an 11-dimensional theory $b$ would be a 2-form field and transform as $\delta_a b_{\mu_1\mu_2} = 2 (\partial_{[\mu_1} a) \phi_{\mu_2]}$.}
\begin{align}
\delta_a A_{\mu_1}^{(1)} = (\partial_{\mu_1} a) \varphi^{(1)} \, , 
\quad
\delta_a A_{\mu_1\mu_2}^{(2)} = 2 (\partial_{[\mu_1} a) \varphi_{\mu_2]}^{(2)} + 2 \alpha (\partial_{[\mu_1} a) \varphi^{(1)} A_{\mu_2]}^{(1)} \, , 
\quad
\delta_a b = 0 \, ,
\quad
\delta_a a = 0 \, .
\end{align}
We compute the field equations of motion by varying the action \eqref{eq:mcs_spst} with respect to $A_{\mu_1}^{(1)}$, $A_{\mu_1\mu_2}^{(2)}$ and $b$. The simplifications necessary to obtain the form of the equations of motion presented here are very similar to the ones discussed in section \ref{sec:m_pst}. The equation of motion for $A_{\mu_1\mu_2}^{(2)}$ is
\begin{align}\label{eq:mcs_pst_eom_A2}
0 = \varepsilon^{\mu_1\ldots \mu_5} \partial_{\mu_3} \left( v_{\mu_4} \mathcal{H}_{\mu_5\mu_6}^{(1)} v^{\mu_6} \right) \, .
\end{align}
Similarly we find the equation of motion for $A_{\mu_1}^{(1)}$
\begin{align}
\begin{aligned}\label{eq:mcs_pst_eom_A1}
0 &= \varepsilon^{\mu_1 \ldots \mu_5} \partial_{\mu_2} \left( v_{\mu_3} \mathcal{H}_{\mu_4\mu_5\mu_6}^{(2)} v^{\mu_6} \right) - 2 \alpha \varepsilon^{\mu_1 \ldots \mu_5} H_{\mu_2\mu_3}^{(1)} \left( v_{\mu_4} \mathcal{H}_{\mu_5\mu_6}^{(1)} v^{\mu_6} \right) \\
&\quad + 2 \alpha \varepsilon^{\mu_1 \ldots \mu_5} A_{\mu_2}^{(1)} \partial_{\mu_3} \left( v_{\mu_4} \mathcal{H}_{\mu_5\mu_6}^{(1)} v^{\mu_6} \right) 
+ \alpha \varepsilon^{\mu_1 \ldots \mu_5} v_{\mu_2} (f_{\mu_6} - (\star_2 f_{\mu_6})) v^{\mu_6} \partial_{\mu_3} (\star G_{\mu_4\mu_5}^{(2)}) \, .
\end{aligned}
\end{align}
Finally, the equation of motion for $b$ is
\begin{align}\label{eq:mcs_pst_eom_b}
0 = \alpha \varepsilon^{\mu_1 \ldots \mu_5} \left( \partial_{\mu_1} (\star G_{\mu_2\mu_3}^{(2)}) \right) \left( v_{\mu_4} \mathcal{H}_{\mu_5\mu_6}^{(1)} v^{\mu_6} - \partial_{\mu_4} ( v_{\mu_5} (f_{\mu_6} - (\star_2 f_{\mu_6})) v^{\mu_6} ) \right) \, .
\end{align}
The first equation is nothing but a gauge transformation which is solved by
\begin{align}
\mathcal{H}_{\mu_1\mu_2}^{(1)} (\partial^{\mu_2} a) = 0 \, .
\end{align}
This then reduces the equation of motion for $b$ to an equation which is solved by
\begin{align}
f_{\mu_1} - (\star_2 f_{\mu_1}) = 0 \, .
\end{align}
Finally, this implies 
\begin{align}
\mathcal{H}_{\mu_1\mu_2\mu_3}^{(2)} (\partial^{\mu_3} a) = 0
\end{align}
as the solution for \eqref{eq:mcs_pst_eom_A1}. These three equations are equivalent to
\begin{align}\label{eq:mcs_pst_eom_sol}
\mathcal{H}_{\mu_1\mu_2}^{(1)} = 0 \, , \quad \mathcal{H}_{\mu_1\mu_2\mu_3}^{(2)} = 0 \, , \quad f_{\mu_1} = (\star_2 \, f_{\mu_1}) \, .
\end{align}
We proceed to compute the string equations of motion and find for $y_{(1)}^{\mu_1}$ using the Dirac vetos
\begin{align}\label{eq:mcs_pst_eom_y1}
0 = e \, \varepsilon^{\mu_1\ldots \mu_5} \partial_{\mu_3} \left( v_{\mu_4} \mathcal{H}_{\mu_5\mu_6}^{(1)}(y_{(1)}) v^{\mu_6} \right) \, .
\end{align}
Similarly we find for $y_{(2)}^{\mu_1}$
\begin{align}\label{eq:mcs_pst_eom_y2}
0 = g \, \varepsilon^{\mu_1 \ldots \mu_5} \, \partial_{\mu_2} \left( v_{\mu_3} \mathcal{H}_{\mu_4\mu_5\mu_6}^{(2)}(y_{(2)}) v^{\mu_6} 
+ \alpha f_{\mu_3}(y_{(2)}) \left( v_{\mu_4} \mathcal{H}_{\mu_5\mu_6}^{(1)}(y_{(2)}) v^{\mu_6} \right) \right) \, .
\end{align}
As expected these equations are solved by \eqref{eq:mcs_pst_eom_sol}. The equations of motion for $z_{(1)}^{\mu_1}$ and $z_{(2)}^{\mu_1}$ are 
\begin{align}\label{eq:mcs_pst_eom_z1}
m_1 \, \partial_J \left( \sqrt{-\gamma_{(1)}} \gamma_{(1)}^{IJ} \partial_I z_{(1) \, \mu_{1}} \right) = \frac{e}{2} F_{\mu_1\mu_2}^{(1)}(z_{(1)}) \frac{\partial z_{(1)}^{\mu_2}}{\partial \sigma_1} 
\end{align}
and
\begin{align}\label{eq:mcs_pst_eom_z2}
\begin{aligned}
m_2 \, \partial_J &\left( \sqrt{-\gamma_{(2)}} \gamma_{(2)}^{IJ} \partial_I z_{(2) \, \mu_{1}} \right) \\
&= \frac{g}{4} \varepsilon_{\mu_1 \ldots \mu_5} \bigg(
 F^{(1) \, \mu_4\mu_5}(z_{(2)}) + 2 \alpha f_{\mu_6}(z_{(2)})F^{(2) \, \mu_4\mu_5\mu_6}(z_{(2)}) \bigg) \frac{\partial z_{(2)}^{\mu_2}}{\partial \sigma_1} \frac{\partial z_{(2)}^{\mu_3}}{\partial \sigma_2} 
\end{aligned}
\end{align}
in agreement with \eqref{eq:mcs_d_eom_z1} and \eqref{eq:mcs_d_eom_z2} when the Dirac vetos and duality relations are taken into account.

\subsubsection*{PST to Dirac}
The Dirac action \eqref{eq:mcs_sd} (with a self dual $f_{M_1}$) is obtained from \eqref{eq:mcs_spst} by setting
\begin{align}
\mathcal{H}_{\mu_1\lambda}^{(1)} v^\lambda = 0 \, , \quad (\star_2 f_{M_1}) = f_{M_1} \, .
\end{align}
The former equation implies
\begin{align}
\begin{aligned}
- \frac{1}{8} v^\rho \mathcal{H}_{\rho\mu_1\mu_2}^{(2)} \mathcal{H}^{(2) \, \mu_1\mu_2\lambda} v_\lambda 
&= - \frac{1}{8} H_{\mu_1\mu_2}^{(1)} H^{(1) \, \mu_1\mu_2} - \frac{1}{24} \varepsilon^{\mu_1\ldots \mu_5} H_{\mu_1\mu_2\mu_3}^{(2)} H_{\mu_4\mu_5}^{(1)} \\
&\quad + \frac{1}{24} H_{\mu_1\mu_2\mu_3}^{(2)} H^{(2) \, \mu_1\mu_2\mu_3} \, .
\end{aligned}
\end{align}
The latter relation removes the last term in \eqref{eq:mcs_pst_Ig}. We proceed as in section \ref{sec:m_pst} and use some integration by parts as well as $(\star G^{(1)}) \wedge (\star G^{(2)}) = 0$ and $(\star G^{(2)}) \wedge (\star G^{(2)}) = 0$ to reach \eqref{eq:mcs_sd} in accordance with figure \ref{fig:diag}.

\subsubsection*{PST to Bunster-Henneaux}
The Bunster-Henneaux action of the 5-dimensional MCS theory is obtained from the PST action \eqref{eq:mcs_spst} by setting $v_{\mu_1} = (1,0,0,0,0)$. This initially gives the action $S_\mathrm{PST}^\mathrm{BH}$ (see \eqref{eq:m_spst_bh}) from which we obtain the canonical momenta. Let $m_i \in \{1,\ldots,4\}$ and find
\begin{align}
\begin{aligned}
&\pi = \frac{\partial \mathcal{L}_\mathrm{PST}^\mathrm{BH}}{\partial (\partial_0 b)} = - \frac{\alpha}{4} \varepsilon^{0m_1 \ldots m_4} (\star G_{m_1m_2}^{(2)}) H_{m_3m_4}^{(1)} + \alpha f_{m_1} J_g^{0m_1} \, , \\
&\pi^{m_1} = \frac{\partial \mathcal{L}_\mathrm{PST}^\mathrm{BH}}{\partial (\partial_0 A_{m_1}^{(1)})} = \frac{1}{9} \varepsilon^{0m_1 \ldots m_4} F_{m_2m_3m_4}^{(2)} + \frac{\alpha}{6} \varepsilon^{0m_1 \ldots m_4} A_{m_2}^{(1)} F_{m_3m_4}^{(1)} + \frac{\alpha}{2} \varepsilon^{0m_1 \ldots m_4} f_{m_2} (\star G_{m_3m_4}^{(2)}) \, , \\
&\pi^{m_1m_2} = \frac{\partial \mathcal{L}_\mathrm{PST}^\mathrm{BH}}{\partial (\partial_0 A_{m_1m_2}^{(2)})} = - \frac{1}{6} \varepsilon^{0m_1 \ldots m_4} F_{m_3m_4}^{(1)} \, .
\end{aligned}
\end{align}
Subsequently, the Hamiltonian is
\begin{align}\label{eq:mcs_pst_bh_h}
\begin{aligned}
\mathcal{H}_\mathrm{PST}^\mathrm{BH} &= f_0 \pi + F_{0m_1}^{(1)} \pi^{m_1} + \frac{3}{2} \partial_{[0} A_{m_1m_2]}^{(2)} \pi^{m_1m_2} - \frac{\alpha}{2} A_0^{(1)} F_{m_1m_2}^{(1)} \pi^{m_1m_2} - \mathcal{L}_\mathrm{PST}^\mathrm{BH} \\
&= \frac{1}{4} H_{m_1m_2}^{(1)} H^{(1) \, m_1m_2} + \frac{1}{12} H_{m_1m_2m_3}^{(2)} H^{(2) \, m_1m_2m_3} + \alpha f_{m_1} (\star f_0) J_g^{0m_1} \\
&\quad - \frac{1}{2} F_{m_1m_2}^{(1)} G^{(1) \, m_1m_2} + \frac{1}{6} F_{m_1m_2m_3}^{(2)} G^{(2) \, m_1m_2m_3} 
+ \frac{\alpha}{2} f_{m_1} F_{m_2m_3}^{(1)} G^{(2) \, m_1m_2m_3} 
\end{aligned}
\end{align}
and some integration by parts yields the canonical action
\begin{align}\label{eq:mcs_pst_can}
S_\mathrm{PST}^\mathrm{can} = \int \mathrm{d}^5x \ \left[ (\partial_0 b) \pi + \partial_0 A_{m_1}^{(1)} \pi^{m_1} + \frac{1}{2} \partial_{0} A_{m_1m_2}^{(2)} \pi^{m_1m_2} - \mathcal{H}_\mathrm{PST}^\mathrm{BH} \right] + I_p \, .
\end{align}
These equations are the second major result of this paper. They extend the results of \cite{Bunster:2011qp} for the MCS theory to include the coupling to branes. We see that all dependencies on $A_0^{(1)}$ and $A_{0m_1}^{(2)}$ have dropped out.\footnote{Note that $(\star_2 \, f_0)$ does not depend on $A_0^{(1)}$.} The canonical action \eqref{eq:mcs_pst_can} and the Hamiltonian \eqref{eq:mcs_pst_bh_h} are of the same general form as their Maxwell counterparts \eqref{eq:m_pst_can} and \eqref{eq:m_pst_bh_h} derived from the Maxwell PST action. However, we cannot convert these expressions into something comparable to \eqref{eq:m_sbh_3} because of the Chern-Simons term in $F_{m_1m_2m_3}^{(1)}$.

Also, the Bunster-Henneaux equations of motion can be obtained from the PST equations of motion \eqref{eq:mcs_pst_eom_A2} - \eqref{eq:mcs_pst_eom_b} by setting $v_{\mu_1} = (1,0,0,0,0)$. The equation of motion for $A_{m_1m_2}^{(2)}$ is 
\begin{align}
0 = \partial_{m_3} \left( H^{(2) \, m_1m_2m_3} + \varepsilon^{0m_1 \ldots m_4} \partial_0 A_{m_4}^{(1)} + G^{(2) \, m_1m_2m_3} \right) \, .
\end{align}
The equation of motion for $b$ is
\begin{align}
0 = \alpha \, \partial_{m_1} (\star G_{m_2m_3}^{(2)}) \left( F^{(2) \, m_1m_2m_3} - \varepsilon^{0 m_1 \ldots m_4} \left( \partial_0 f_{m_4} - \partial_{m_4} (\star f_0) \right) \right) \, .
\end{align}
Finally, taking the other two equations into account, the equation of motion for $A_{m_1}^{(1)}$ is
\begin{align}
\begin{aligned}
0 &= \alpha F_{m_2m_3}^{(1)} H^{(2) \, m_1m_2m_3} - \alpha A_{m_2}^{(1)} \partial_{m_3} H^{(2) \, m_1m_2m_3} - \partial_{m_2} H^{(1) \, m_1m_2} \\
&\quad - \frac{1}{2} \varepsilon^{0 m_1 \ldots m_4} \left( \partial_0 \partial_{m_2} A_{m_3m_4}^{(2)} - \alpha \partial_0 A_{m_2}^{(1)} F_{m_4m_4}^{(1)} + \alpha F_{m_2m_3}^{(1)} (\star G_{0m_4}^{(2)}) \right) \\
&\quad - \frac{1}{2} \varepsilon^{0 m_1 \ldots m_4} \partial_{m_2} \left( (\star G_{0m_3m_4}^{(1)}) + 2 \alpha (\star f_0) (\star G_{m_3m_4}^{(2)}) - 2 \alpha \left( A_{m_3}^{(1)} + 2 f_{m_3} \right) (\star G_{0m_4}^{(2)}) \right) \, . 
\end{aligned}
\end{align}
These equations are solved exactly as in the Maxwell case and we once again obtain \eqref{eq:mcs_pst_eom_sol}. Similarly, the string equations of motion follow from the PST string equations. 

\section{Outlook and Conclusion}\label{sec:conc}
In this paper, we studied three approaches to electric-magnetic duality in the 4-dimensional Maxwell theory coupled to a dyonic point charge and the 5-dimensional Maxwell-Chern-Simons theory coupled to an electric point charge and a magnetic string charge. The main results are the PST and Bunster-Henneaux formulations of the 5-dimensional MCS theory presented in section \ref{sec:mcs_pst}. In particular, we showed that our PST formulation consistently reduces to the Dirac formulation of \cite{Bekaert:2002eq}. It appears that this is not the case for the PST formulation of 11-dimensional supergravity coupled to the M2- and M5-brane in \cite{Bandos:1997gd}. However, generalizing our result to 11-dimensions provides an easy fix.

A natural next step in this line of work is to consider $p$-form electrodynamics with a quadratic Chern-Simons interaction term. The simplest case is that of a 1-form gauge field and a 0-form dual field in 3-dimensional Minkowski space. However, because the number of spacetime dimensions is so small there is no room for an auxiliary field $b$ on any of the branes. Hence one must solve the equation corresponding to \eqref{eq:mcs_eom_ext_d} solely by imposing boundary conditions such as in \cite{Henneaux:1986tt}. But without an auxiliary field the introduction of a dual field strength as described in the beginning of section \ref{sec:mcs_pst} is difficult. Thus one must consider a higher dimensional theory with a quadratic Chern-Simons interaction term such that an auxiliary field can live on one of the branes. Without sources the twisted self duality of 1-form electrodynamics with a quadratic Chern-Simons interaction term in 3 spacetime dimensions has been discussed in \cite{Bunster:2011qp}.

\section{Acknowledgments}
I would like to thank Axel Kleinschmidt and Hermann Nicolai for many helpful discussions and Oleg Evnin, Axel Kleinschmidt, Karapet Mkrtchyan and Hermann Nicolai for their comments on the manuscript.

This project has received funding from the European Research Council (ERC) under the European Union’s Horizon 2020 research and innovation programme (grant agreement no. 740209).

\end{document}